\documentclass[
    prapplied,
    preprint,
    reprint,
    superscriptaddress,
    ]{revtex4-2}

\usepackage{amsmath,amssymb}
\usepackage{graphicx}
\graphicspath{{figures/}}
\usepackage{dcolumn}
\usepackage{bm}
\usepackage{ulem}
\usepackage{soul}
\usepackage[utf8]{inputenc}
\usepackage[T1]{fontenc}
\usepackage{mathptmx}
\usepackage{etoolbox}
\usepackage{siunitx}
\DeclareSIUnit{\belmilliwatt}{Bm}
\DeclareSIUnit{\dBm}{\deci\belmilliwatt}
\DeclareUnicodeCharacter{0300}{}
\DeclareUnicodeCharacter{0301}{}
\DeclareUnicodeCharacter{0327}{}

\newcommand{\figref}[1]{Figure~\ref{#1}}

\usepackage{color,soul} 
\definecolor{darkblue}{rgb}{0,0,0.7}
\definecolor{darkgreen}{rgb}{0.2,0.95,0.2}
\definecolor{darkred}{rgb}{.7,0,0}
\definecolor{purple}{rgb}{0.7,0,0.9}
\definecolor{orange}{rgb}{1,0.5,0}
\definecolor{grey}{rgb}{0.6,0.6,0.6}
\definecolor{lightpink}{rgb}{1,0.7,0.75}
\definecolor{pink}{rgb}{1,0.4,0.58}

\begin{document}

\title{Ultrashort electron wavepackets via frequency-comb synthesis}

\author{Matteo Aluffi}
\thanks{These authors contributed equally to this work.}
\affiliation{Universit\'e Grenoble Alpes, CNRS, Grenoble INP, Institut N\'eel, 38000 Grenoble, France}

\author{Thomas Vasselon}
\thanks{These authors contributed equally to this work.}
\affiliation{Universit\'e Grenoble Alpes, CNRS, Grenoble INP, Institut N\'eel, 38000 Grenoble, France}

\author{Seddik Ouacel}
\affiliation{Universit\'e Grenoble Alpes, CNRS, Grenoble INP, Institut N\'eel, 38000 Grenoble, France}

\author{Hermann Edlbauer}
\affiliation{Universit\'e Grenoble Alpes, CNRS, Grenoble INP, Institut N\'eel, 38000 Grenoble, France}

\author{Cl\'ement Geffroy}
\affiliation{Universit\'e Grenoble Alpes, CNRS, Grenoble INP, Institut N\'eel, 38000 Grenoble, France}

\author{Preden Roulleau}
\affiliation{Universit\'e Paris-Saclay, CEA, CNRS, Gif sur Yvette, 91191, France}

\author{D. Christian Glattli}
\affiliation{Universit\'e Paris-Saclay, CEA, CNRS, Gif sur Yvette, 91191, France}

\author{Giorgos Georgiou}
\affiliation{James Watt School of Engineering, Electronics and Nanoscale Engineering, \\ University of Glasgow, Glasgow, G12 8QQ, United Kingdom}

\author{Christopher B\"auerle}
\thanks{Corresponding Author: christopher.bauerle@neel.cnrs.fr}
\affiliation{Universit\'e Grenoble Alpes, CNRS, Grenoble INP, Institut N\'eel, 38000 Grenoble, France}

\begin{abstract}
Single-electron sources are an essential component of modern quantum nanoelectronic devices.
Owing to their high accuracy and stability, they have been successfully employed for metrology applications, studying fundamental matter interactions and more recently for electron quantum optics.
They are traditionally driven by state-of-the-art arbitrary waveform generators that are capable of producing single-electron pulses in the sub-\SI{100}{\ps} timescale.
In this work, we use an alternative approach for generating ultrashort electron wavepackets.
By combining several harmonics provided by a frequency comb, we synthesise Lorentzian voltage pulses and then use them to generate electron wavepackets.
Through this technique, we report on the generation and detection of an electron wavepacket with temporal duration of \SI{27}{\ps} generated on top of the Fermi sea of a 2-dimensional electron gas -- the shortest reported to date.
Electron pulses this short enable studies on elusive, ultrafast fundamental quantum dynamics in nanoelectronic systems and pave the way to implement flying electron qubits by means of Levitons.

\end{abstract}

\maketitle


\section*{Introduction}
Electronic excitations in solid-state, nanoelectronic systems provide a unique platform for investigating fundamental matter interactions, such as many-body interactions, the fractional quantum Hall effect, Fermion statistics as well as Abelian and non-Abelian statistics \cite{Stern2008, Oliver1999, Henny1999, Bocquillon2013, Dubois2013, Freulon2015, Kapfer2019, Bartolomei2020, Nakamura2020, Taktak2022}.
The ability to control such excitations at the individual electron level opens exciting perspectives in terms of quantum information processing and real-time control of topologically protected excitations \cite{Bocquillon2013, Taktak2022, Kamata2022}.
In this respect, the advent of highly efficient and precise single-electron sources was found to be technologically very important for metrology and more specifically for the definition of the quantum current standard \cite{Giblin2020, Scherer2019, Kaneko2016}. 
In addition to metrology applications, the quality of these sources has made them very popular in the field of quantum nanoelectronics.
Over the last few years they have been the driving force for the development of electron quantum optics, where -- in analogy to quantum optics -- quantum information is encoded in the charge
or spin degree of freedom of single propagating electrons \cite{Baeuerle2018, Edlbauer2022, Takada2019, Wang2022, Wang2022a, Ubbelohde2022, Fletcher2022, Jadot2021}.

There are currently several types of single-electron sources, with the majority of them being based on isolating electrons in well-defined regions through electrostatic barriers, that can accommodate single or few electrons, 
which are then injected into nanoelectronic circuits \cite{Feve2007,Blumenthal2007,Hermelin2011,McNeil2011, Wang2022a}.
For such single electron sources one has to distinguish the following two cases: (i) The injected electrons have energies of several tens meV, well separated fromm the bulk electrons. As a consequence, the electron-electron interaction with other electrons in the Fermi sea is suppressed.
On the other hand, this makes these systems more prone to decoherence due to fluctuations in the electromagnetic environment such as dopants \cite{Wang2022a} or due to electron-phonon interactions \cite{emary2019energy, Clark2020}. 
The main application for these kind of quantum-dot-based sources is in the field of metrology, where current advances report electron pumping accuracy of as high as \SI{0.92}{ppm} (parts per million) \cite{Yamahata2016}.
In addition to metrology, their use for electron quantum optics is still an open research topic \cite{Takada2019, Wang2022, Edlbauer2022, Wang2022a, Fletcher2022,Ubbelohde2022}.
(ii) The electron is injected from a larger quantum dot into the edge channel of a quantum Hall system \cite{Feve2007}. 
As a consequence, the electrons are emitted much closer to the Fermi sea, typically in the tens of $\mu$eV range. 
As the bandwidth of the wavepacket is smaller than the energy of emission they are still well separated from the Fermi sea and coherent electron transport has already been observed \cite{Bocquillon2013}.

An alternative technique for generating single electrons relies on temporally short voltage pulses, also known as Levitons.
Typically, any voltage signal that is applied on an electron reservoir can induce the motion of free carriers (electrons and holes). 
It was however postulated by Levitov {\it et al.} \cite{Levitov1996, Keeling2006}, that voltage pulses with Lorentzian profile enable pure electron excitation without any holes.
The amount of excited electrons can be controlled by the amplitude of the pulses and can be tuned such that only a single electron is excited from the reservoir.
In contrast to quantum-dot-based sources, Levitons are excited very close to the Fermi sea, they do not entangle with their environment and can be considered as classical current pulses \cite{Ferraro2014}. 
The first successfully implemented Leviton source \cite{Dubois2013} enabled experiments on electron tomography \cite{Jullien2014, Bisognin2019} and time-resolved reconstruction of electron wavefunctions \cite{Roussely2018}.
In addition, recent theoretical works predict that Levitons can be used to study entanglement
and non-locality in Mach-Zehnder interferometers \cite{Vyshnevyy2013, Dasenbrook2016a} or to generate an effective fractional charge e/2 \cite{Moskalets2016}.
In the form of ultrashort wavepackets, Levitons are also discussed as promising approach to generate electron flying qubits \cite{Baeuerle2018,Edlbauer2022}.

The generation of Levitons has been mainly driven by state-of-the-art arbitrary waveform generators (AWGs) that have the capacity of generating single-electron pulses with temporal duration as short as \SI{42}{\ps} \cite{Bisognin2019}.
Although using an AWG for generating single-electron pulses is a powerful technique, 
one cannot create very short Lorentzian voltage pulses with 
sufficient
precision due to limited bandwidth and sampling rate.
These AWG limitations degrade the signal quality by generating unwanted  hole excitations. 
With the shortest timescales provided by state-of-the-art AWGs, access to elusive dynamics of low-dimensional systems -- such as crystallisation and fractionalisation \cite{Ronetti2018}, dynamically control of interference patterns \cite{Gaury2014} and spectroscopy of quantum-coherent circuits \cite{Burset2019} -- is so far not possible.

An alternative to the generation of Levitons through AWGs is to synthesise Lorentzian-shaped voltage pulses using frequency combs.
The discovery of frequency combs has revolutionised optical spectroscopy and metrology in the visible and ultraviolet region of the electromagnetic spectrum \cite{Udem2002}.
Although very popular in optics, frequency combs have not been utilised extensively for radio-frequencies (RF).

In this paper, we use RF frequency combs to synthesize ultrashort Lorentzian voltage pulses that enable stable generation of single-electron Levitons.
By controlling the amplitude and phase of each harmonic of the comb, we show the capability to finely tune the extent of the charge excitation enabling Leviton formation 
and -- more importantly -- the precise correction of dispersion from the coaxial line bringing the voltage pulse to the sample.
Additionally, we demonstrate that the temporal duration of Levitons is tuneable from the continuous limit (sinusoidal waves) down to electron wavepackets with \SI{20}{\ps} duration.
Finally, we inject the output from our frequency-comb generate into a quantum nanoelectronic device and perform a time-resolved measurement of synthesized Lorentzian voltage pulses
{in-situ} on a cryogenic setup.
This work sets strong foundations for exploring time-resolved ultrafast electron dynamics in solid-state systems and opens possibilities for a competitive electron flying qubit technology.


\section*{Arbitrary voltage pulse generation}
\begin{figure}[ht!]
\includegraphics[scale=1]{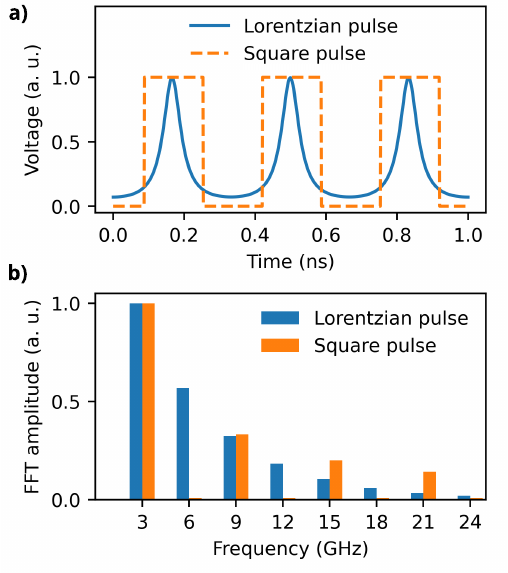}
\caption{
Principle of Fourier synthesis.
(a) and (b) The principle of Fourier synthesis explained through a Lorentzian (blue) and square (orange) voltage pulses. A periodic pulse (a) in the time domain can be decomposed into a series of harmonics in the frequency domain (b), whose amplitude and phase (not shown here) varies as a function of frequency.}

\label{fig:fourier_synth}
\end{figure}

\begin{figure}[ht!]
\includegraphics[scale=1]{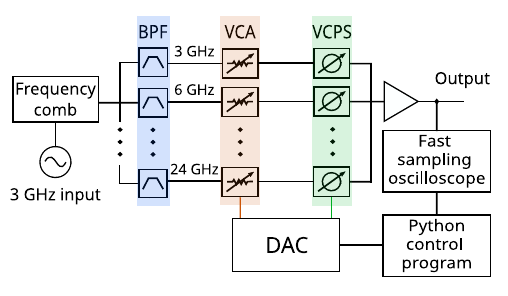}
\caption{
The arbitrary voltage pulse synthesiser block diagram. 
A 3 \rm{GHz} fundamental frequency is fed through a commercially available frequency comb, which generates higher harmonics. 
Each harmonic is then selected by a band-pass filter (BPF), attenuated by a voltage-controlled attenuator (VCA) and delayed by the voltage-controlled phase-shifter (VCPS).
The output is measured in the time-domain with a sampling oscilloscope, which provides feedback to a homemade  control software to automatically tune the VCAs and VCPSs using a programmable digital to analog converter (DAC).
}
\label{fig:block_diagram}
\end{figure}


Any periodic signal can be represented by a Fourier series, which is a linear superposition of orthonormal functions, such as sinus waves.
As shown in \figref{fig:fourier_synth}~(a) and~(b), a square pulse (orange curve) is for instance described by a series of odd harmonics with a 1/n decaying amplitude as a function of frequency.
In the same manner, a Lorentzian pulse (blue curve) can be decomposed into a series of harmonics whose amplitude falls exponentially.
The frequency-comb synthesizer that we demonstrate in this work is now a precise experimental implementation of this mathematical concept.

\begin{figure}[t!]
\includegraphics[scale=1]{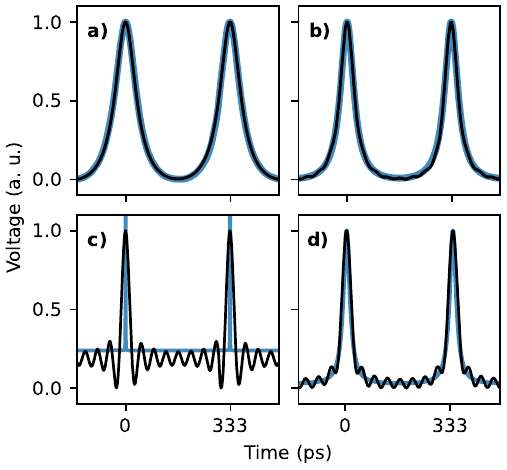}
\caption{
Voltage pulses generated by the frequency comb synthesiser. The black solid line corresponds to the pulse measured with a fast oscilloscope, and the blue line to the theoretical voltage pulse fed into the optimisation algorithm.
(a) and (b) are Lorentzian voltage pulses with duration \SI{80}{\ps} and \SI{50}{\ps}, respectively.
(c) Impulse response of the synthesiser, where the input to the optimisation algorithm is a delta-like function and 
(d) \SI{25}{\ps} Lorentzian voltage pulse used for the measurement in \figref{fig:cryo_experiment}~(b).}
\label{fig:Lorentzian_width}
\end{figure}

Multiple harmonic microwave signals with proper adjustment of amplitude and phase are mixed to obtain a Lorentzian target pulse -- the central ingredient for Leviton formation.
As illustrated by the block diagram of \figref{fig:block_diagram}, a fundamental sinusoidal signal with a frequency of \SI{3}{\GHz} and a power of \SI{19}{\dBm} is fed into a frequency comb of the type MARKI NLTL-6026. 
This microwave component is a Schottky-diode-based transmission line with the ability to generate several harmonics from any fundamental frequency up to \SI{50}{\GHz}.
After generation, the harmonics are 
spread into different paths
by an eight-way RF splitter.
A \SI{200}{\MHz} bandwidth band-pass filter selects 
for each path a specific harmonic frequency.
The power of each harmonic signal is determined by the selection of microwave components, such as splitters or band-pass filters, giving for the current implementation \SI{-22.8}{\dB} at \SI{3}{\GHz} to \SI{-53.1}{\dB} at \SI{24}{\GHz}.
After filtering, each harmonic passes through a voltage-controlled attenuator (VCA) and a voltage-controlled phase-shifter (VCPS) that automatically adjust the amplitude and phase 
for each harmonic signal.
Having set the desired harmonic composition, the recombined signal provides the targeted voltage pulse on the output that is finally
amplified with a Marki AMZ-40 and measured with a sampling oscilloscope.

The automatic adjustment of the amplitudes and phases is accomplished by a feedback-loop.
As shown in \figref{fig:block_diagram}, the output of our frequency-comb synthesizer is monitored by a digital sampling oscilloscope. 
The continuously captured data is Fourier transformed, and the amplitudes and phases of each harmonic are extracted and compared with the values of the desired pulse. 
The Python control program then adjusts the voltage applied to the VCAs and VCPSs by means of a dichotomy algorithm.
After typically 20 iterations an almost perfect waveform is obtained.
For a Lorentzian-shape voltage pulse, the optimisation algorithm usually converges within a couple of minutes.
It is worthwhile to note that adjusting the attenuators and phase-shifters is only needed once and that, as it is shown below, the generated pulses are stable over days.

In order to generate Leviton pulses in a quantum nanoeletronic device, it is necessary to form Lorentzian charge excitation that carry exactly one or multiple electrons.
Accordingly, the voltage pulse generation must allow for amplitude control that is extremely precise. 
\figref{fig:Lorentzian_width} shows real-time oscilloscope data (black curves) from various synthesised Lorentzian pulses.
The traces of \figref{fig:Lorentzian_width}~(a) and~(b) show the output for target pulses with duration of \SI{80}{\ps} and \SI{50}{\ps}.
Remarkably, they are nearly identical to the theoretical target (blue lines) and have minimum distortion, ensuring generating Levitons with an extremely low number of hole excitations.
The quality of the pulse at zero temperature can be estimated by computing the excess hole excitation, defined as $\frac{Nh}{Ne}$, where $N_h$ and $N_e$ are the numbers of excited holes and electrons respectively (see appendix B). For the \SI{80}{ps} pulse, a relative hole excitation as low as \SI{0.07}{\%} is expected.  
It is worthwhile to note that voltage pulses with similar duration, as the one illustrated here (\figref{fig:Lorentzian_width}~(a) and~(b)), have been used in the past to generate ultrafast Levitons, using state-of-the-art AWGs \cite{Roussely2018, Bisognin2019}. 
Those voltage pulses resembled more a gaussian-like pulse rather than the Lorentzian shape required for the generation of Leviton pulses as demonstrated in Ref.~\cite{Roussely2018}.

Analog Fourier synthesis -- as demonstrated in this work -- enables on the other hand the generation of nearly ideal ultrashort Lorentzian pulses that go beyond what state-of-the-art AWGs can achieve.
To characterise the limit of our device, we have programmed the synthesiser to output a Dirac delta-like pulse (Lorentzian pulse with a width of \SI{0.1}{\ps}). 
This allows to measure the impulse response of the system, which can be used to predict the output waveform for any requested input. 
The output can then be simply computed as a convolution product between the requested input and the impulse response. 
In our case, the impulse response, shown in \figref{fig:Lorentzian_width}~(c), is a cardinal sinus with a Full Width at Half Maximum (FWHM) of \SI{22}{\ps}. 

\figref{fig:Lorentzian_width}~(d) illustrates a \SI{25}{\ps} Lorentzian pulse, which appears to be almost identical to its theoretical line shape (blue curve) and with minimal distortion. 
Even for this case, the generated hole excitations due to imperfect Lorentzian line shape is less than \SI{0.55}{\percent} (see appendix B).
The slight distortion observed to either side of the main peak is due to the absence of harmonics higher than \SI{24}{\GHz}. 
It is worth noting that any limitations imposed by the cutoff frequency of standard SMA connectors used in our setup as well as any dispersion due to the RF cables is automatically compensated by our optimisation algorithm.

\begin{figure}[t!]
\centering
    
\includegraphics[scale=1.0]{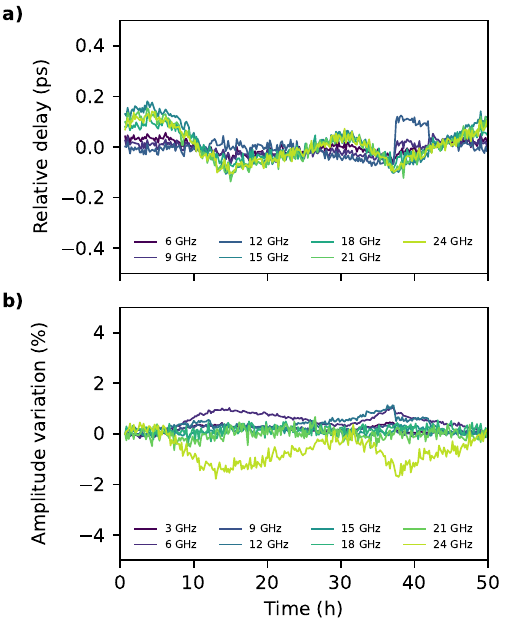}
\caption{ Output stability over time. The synthesiser output was monitored every 10 minutes for 50 hours. (a) Relative time drift of each harmonic, referenced to the \SI{3}{\GHz} fundamental frequency.
(b) Amplitude variation for each harmonic.}
\label{fig:stability}
\end{figure}

One of the main advantages of using a frequency comb based generator, as opposed to individual phase locked frequency generators \cite{Dubois2013},
is that all the harmonics are generated from the same source signal. 
Even if the phase of the fundamental frequency shifts over time, the phases of all higher harmonics will shift in the same manner, and therefore the generated signal will not be distorted.
The stability of the phase between the harmonics of the synthesiser, as well as their amplitude, is of paramount importance as any phase or amplitude variation can distort the shape and duration of the generated pulses.
This means that for a frequency-comb based system, an ultra-stable picosecond-duration pulse can be synthesised once at the beginning of an experiment and then be kept the same over several days. 

The stability of this setup is apparent from the phase and amplitude variation of each harmonic, as shown in \figref{fig:stability}. 
To generate these data, a Lorentzian pulse was measured through a fast oscilloscope, once every ten minutes for 50 hours.
Then, the time-domain data were Fourier transformed, in order to extract the phase and amplitude information for each harmonic. 
\figref{fig:stability}~(a), shows the phase stability of this synthesiser. In order to have a more meaningful representation of the data, \figref{fig:stability}~(a) represents the relative time delay of each harmonic with respect to the master frequency (\SI{3}{\GHz}).
As all higher harmonics are generated from the same master frequency, their relative delay depends only on the passive components used in our setup, such as the filters, attenuators and phase-shifters. 
The $\sim$~\SI{0.2}{\ps} delay variation between the harmonics is mainly due to temperature variations in the laboratory over the course of a 2-day period.
This relative delay is 2 orders of magnitude smaller than the shortest Lorentzian pulse (\SI{25}{\ps}) and can be regarded as negligible.

The amplitude variation of each harmonic is also very stable, as illustrated in \figref{fig:stability}~(b). 
The amplitude of each harmonic, besides being important for determining the duration and shape of the final pulse, 
is also crucial for defining the overall amplitude of the Leviton pulses.
In order to generate single-electron Levitons, it is important to control the amplitude of the generated pulse down to the microvolt level, where indicatively \SI{1}{\uV} corresponds roughly to the addition of \SI{1}{\percent} of an extra electron.
In this setup, the observed amplitude variation is less than \SI{1}{\percent} over the 50-hour acquisition time, and therefore the generated Levitons can contain a single electron with a very small uncertainty.


\section*{Generation of electron wavefunctions in a quantum conductor}

\begin{figure*}[t]
\includegraphics[scale=1]{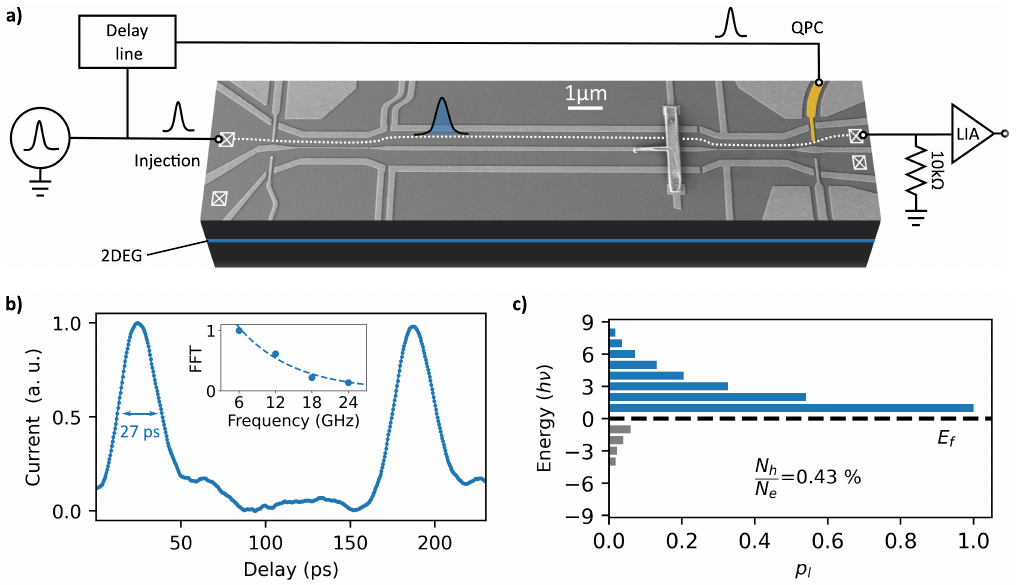}
\caption{Time-resolved measurement of an electron wavepacket using the frequency comb synthesiser.
(a) Scanning tunneling micrograph of the quantum device made from a GaAs/AlGaAs heterostructure with mobility  $\mu_{\rm{e}} = \SI{1.77e6}{\square\cm\per\V\per\s}$ and electron concentration  $\rm{n_{e}} = \SI{1.92e11}{\per\square\cm}$.
The voltage pulse signal generated with the frequency comb synthesiser is split into two. 
One voltage pulse is sent to the upper left Ohmic contact to launch the electron wavepacket while the second voltage pulse is sent to the QPC (highlighted in yellow) via a mechanical delay line. 
The QPC is operated as an ON/OFF switch, following the procedure detailed in \cite{Roussely2018}.
Changing the time delay between the two voltage pulses allows reconstructing in a time-resolved manner the waveform of the generated electron wavepacket.   
The generated current is detected through a \SI{10}{\kohm} resistor, amplified at room temperature and measured by a lock-in amplifier (LIA) using a demodulation technique.
(b) Time resolved measurement of a Leviton pulse. 
The inset shows the Fourier transform of the pulse, which follows an exponential trend and therefore confirming the Lorenztian nature of the generated electron wavepacket.
(c) Calculated excitation spectrum of the electron wavepacket generated in (b). $\lvert p_l \rvert^{2}$ denotes the probability of an electron to absorb or emit $l$-photons of an energy quanta $h\nu$ (see appendix B).
}
\label{fig:cryo_experiment}
\end{figure*}

The generation of single-electron Leviton pulses relies on voltage pulses with Lorentzian
shape, applied directly on the Fermi sea of a quantum conductor.
By perturbing the Fermi sea with these pulses we can generate pure electronic excitations, without any holes accompanying the electrons \cite{Keeling2006, Dubois2013, Jullien2014, Bisognin2019},
in stark contrast to other popular electron sources \cite{Feve2007,Blumenthal2007}.
This unique property of Levitons makes them ideal candidates for electron flying qubit technologies \cite{Edlbauer2022}, as well as
an attractive platform for studying fundamental time-resolved electron-electron interactions.

To showcase the full potential of our novel voltage pulse synthesiser, we demonstrate the generation and time-resolved detection of an electron wavepacket with a record duration of \SI{27}{\ps} in a two-dimensional electron gas (2DEG).
A dilution refrigerator with a base temperature of \SI{20}{\milli\K} was used to cool down our 2DEG to cryogenic temperatures. 
The quantum device consists of a Mach-Zehnder interferometer realized by depositing electrostatic surface gates on top of a GaAs/AlGaAs heterostructure as shown in \figref{fig:cryo_experiment}~(a). 

Applying a set of negative gate voltages to these electrostatic gates allows engineering of the trajectory of the propagating electron wavepacket. 
For the time-resolved measurements only the upper path of the interferometer is used, that means the gate voltages on the central island as well as the adjacent split wires are sufficiently negative such that the electron wavepacket follows the path indicated by the white dashed line.

To perform time-resolved measurements and reconstruct the electron waveform in the time domain, we split the voltage pulse output of our synthesiser into two parts using a power divider.
The first part, denoted as ``Injection'' on \figref{fig:cryo_experiment}~(a), is sent through a computer controlled delay line and then into the injection Ohmic contact of our sample (left white square box in \figref{fig:cryo_experiment}~(a)).
The second part, denoted as ``QPC'', is directly fed into the RF line, which is connected to a Quantum Point Contact (QPC) positioned along the electron wave guide. 
Both, the injection Ohmic contact and the sampling QPC line is equipped with a high bandwidth Bias Tee SHF BT45R in order to be able to apply a DC and AC component. 
For detecting the electrons, we measure the voltage generated across a \SI{10}{\kohm} resistor, which is then amplified and measured by a LockIn Amplifier (LIA).
To enable LIA measurements, the ``master'' source of the synthesiser is modulated at \SI{11.95}{\kHz}.

The QPC line is employed to realise the time-resolved measurement.
In line with Roussely {\it et al.} \cite{Roussely2018}, to reconstruct the time-domain profile of an electron wavepacket, the QPC is negatively biased with a sufficient DC voltage, such that it blocks all transmission through the waveguide.
Then, on top of that DC bias we apply an ultrashort voltage pulse to the 
QPC,
time-delayed by $\Delta\tau$ with respect to the injection pulse.
This voltage pulse will effectively act as a switch, enabling transmission for a very short amount of time.
By appropriately choosing the negative DC bias, the switching time can be much shorter than the duration of the electron wavepacket, therefore enabling accurate reconstruction. 

\figref{fig:cryo_experiment}~(b) is the main result of this paper, showing the shortest ever measured electron wavepacket propagating on top of the Fermi sea with \SI{27}{\ps} duration. 
In order to benchmark the dispersion of our system, the pulse was generated with a relatively high voltage amplitude and contains around $\sim$~20 electrons. The dispersion appears to be minimal, as the injected pulse (Figure \ref{fig:Lorentzian_width} (d)) had a similar width of \SI{25}{ps}.
To better demonstrate the time-resolved measurement technique, we have used a \SI{6}{\GHz} (\SI{166}{\ps}) fundamental frequency for the synthesiser.
This allowed us to measure two pulses within the same scanning window, since the scanning range of our delay line is limited to \SI{230}{\ps}. 
The Fourier transform of the generated electron wavepacket, shown in the inset of \figref{fig:cryo_experiment}~(b)), exhibits an exponential behaviour, which is distinctive of a Lorentzian electron excitation.

Finally, we analyze the amount of excess hole excitation of the generated Lorentzian voltage pulses as well as the generated electron wavepackets by calculating their excitation spectrum at zero temperature in the framework of Floquet scattering theory \cite{Moskalets2002, Dubois2013a} (see Appendix B).
The excitation spectrum for the experimentally measured electron wavepacket of \SI{27}{\ps} time duration is shown in \figref{fig:cryo_experiment}~(c)).
We observe an excess hole excitation that is less than \SI{0.43}{\percent}.
While this is a remarkable result, one needs to bear in mind that this number has been computed at zero temperature.
The excess hole excitation has been experimentally measured in \cite{Bisognin2019} for similarly-generated Lorentzian pulses at \SI{50}{mK} and was found to be 3 \%. Since this value is mostly linked to thermal excitations in the Fermi sea, we expect our device to perform similarly at finite temperatures.
Similar analysis (see appendix B) for the generated voltage pulses display an extremely clean excitation spectrum with less than \SI{0.22}{\percent}
hole excitations when the voltage pulses have a duration of \SI{50}{\ps} or more. 
For the shortest voltage pulses (\SI{25}{\ps}), we obtain a maximum of \SI{0.55}{\percent}
hole excitations. 
Adding additional harmonics in future will further allow improving the signal quality as well as shortening the generated voltage pulses.
With a fundamental frequency of 3 or 6 GHz, the maximum of harmonic frequencies lies at \SI{36}{\GHz} due to limited frequency bandwidth of the radio-frequency components.
Within this range, it is possible to generate voltage pulses of a duration down to \SI{14.6}{\ps}.
It is possible, on the other hand, to create even shorter voltage pulses, in the range of \SI{1}{\ps}, using an optoelectronic approach \cite{Auston1975}, but unfortunately their integration at cryogenic temperatures \cite{Georgiou2020} still remains a challenge.\\

To conclude, in this paper we have demonstrated a novel technique for generating 
Levitions (single-electron Lorentzian electron wavepackets)
based on a frequency-comb synthesiser.
The
synthesized Levitons are the shortest ever generated electron wavepackets riding on the Fermi sea, with a record duration of \SI{27}{\ps}. 
We evaluate the spurious hole contribution to be below \SI{0.43}{\percent} which make these Levitons ideal carriers of quantum information in the field of electron quantum optics, as well as for studying fundamental matter interactions in the time domain.


\begin{figure*}
\includegraphics[scale=1]{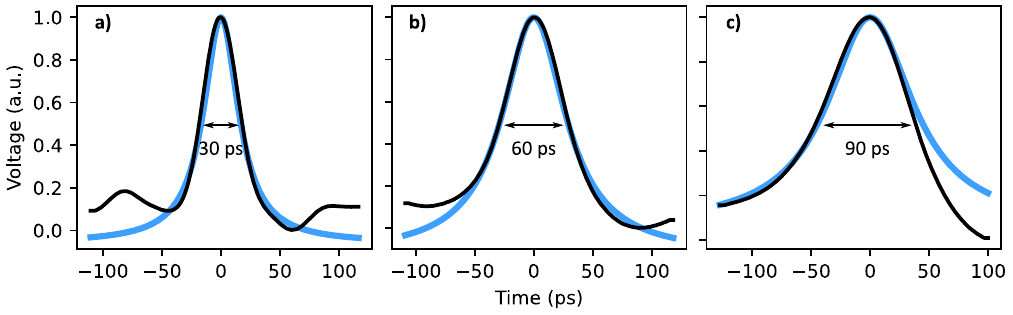}
\caption{Synthesised electron wavepackets with varying duration, (a)~\SI{30}{\ps}, (b)~\SI{60}{\ps} and~(c)~\SI{90}{\ps}, measured through the 2DEG of the sample shown in Figure 5 of the main text. The black lines are experimental data and the blue lines correspond to the best Lorentzian fit.}
\label{suppfig:varying_width}
\end{figure*}


\section*{Acknowledgements}

This project has received funding from the European Union’s H2020 research and innovation programme under grant agreement No 862683, ‘UltraFastNano’.
C. B. and D. C. G. acknowledge funding from the French Agence Nationale de la Recherche (ANR), Project Fully Quantum ANR-16-CE30-0015. 
C.B. acknowledges funding from the French Agence Nationale de la Recherche (ANR), Project ANR QCONTROL ANR-18-JSTQ-0001.
C.B., D.C.G., and P.R. acknowledge funding from the French Agence Nationale de la Recherche (ANR), Programmes et Équipements Prioritaire de Recherche (PEPR) Technologies Quantiques, Project E-QUBIT-FLY, ANR-22-PETQ-0012.
T. V. acknowledges funding from the French Laboratory of Excellence project "LANEF" (ANR-10-LABX-0051).
M.A. acknowledges the MSCA co-fund QuanG Grant No. 101081458, funded by the European Union.\\
\indent Views and opinions expressed are those of the author(s) only and do not necessarily reﬂect those of the European Union or the granting authority. Neither the European Union nor the granting authority can be held responsible for them.

\section*{Appendix A: Synthesised electron wavepackets}

\begin{figure}
\includegraphics[scale=1]{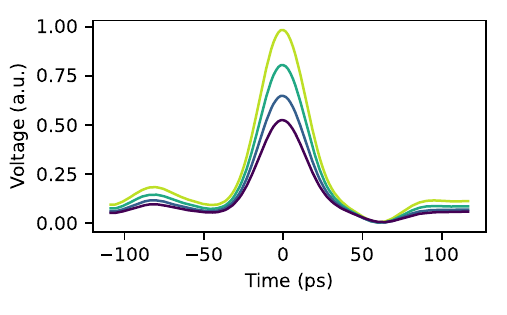}
\caption{Synthesised lorentzian voltage pulses with \SI{30}{\ps} duration, measured through the 2DEG of the sample, shown in Figure 5 of the main text. The amplitude of the pulses is varied continuously.}
\label{suppfig:varying_amplitude}
\end{figure}

To demonstrate the feasibility of generating wavepackets containing an arbitrary number of electrons and of any duration, we have performed Fourier synthesis of various Lorentzian pulses with different amplitudes and widths, shown in Figures~\ref{suppfig:varying_width} and~\ref{suppfig:varying_amplitude}. 
As can be appreciated from the data, the proposed pulse generator is very versatile,  as it allows for the generation of extremely high quality voltage pulses of any temporal duration and amplitude. 
As described in the main results of the paper, this is due to the independent tuneability of each harmonic and the extreme stability between the frequency comb harmonics (see Figures 2 and 4).


One of the greatest challenges of transmitting ultrashort pulses over long radio-frequency (RF) cables is dispersion.
The effect of dispersion is unfortunately more prominent with ultrashort pulses that exhibit a broad spectrum and can result into pulse broadening and chirping (deformation). 
To be able to generate perfect lorentzian pulses at cryogenic temperatures (mK), it is important to properly calibrate and pre-compensate for any dispersion introduced by the transmission lines. 
By feeding the scattering parameters of the transmission lines into our automated algorithm, we can compensate for the dispersion introduced to our pulse. 
With this approach we can automatically take into account for dispersion owning to an arbitrary length of transmission line, therefore having perfect lorentzian pulses at the sample stage.
Both Figures~\ref{suppfig:varying_width} and~\ref{suppfig:varying_amplitude} are time-resolved measurements performed through the 2DEG sample, with the method explained in Figure 5 of the main text.


\begin{figure*}
    \centering
    \includegraphics[scale=1]{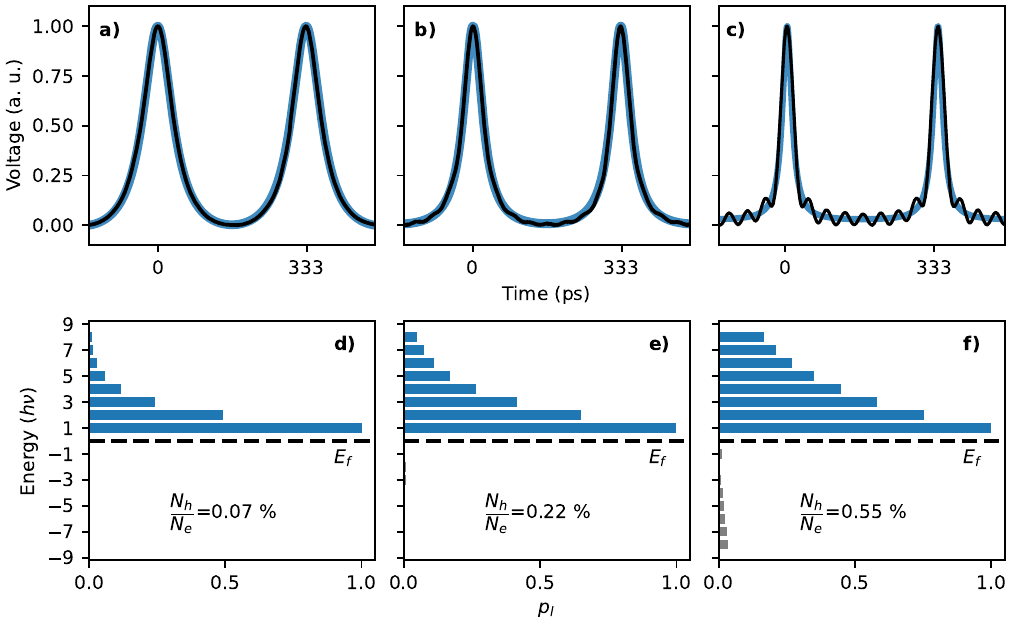}
    \caption{Excitation spectra of the signal generator output pulses. (a), (b) and (c) measured Lorentzian voltage pulses (black) at the signal generator output with their corresponding Lorentzian fit (blue). The FWHMs are \SI{80}{ps}, \SI{50}{ps} and \SI{25}{ps} respectively. (d), (e) and (f) expected excitation spectra for electron wave packets with same waveform and integer charge per pulse. }
    \label{suppfig:pulses_spectrum}
\end{figure*}

\section*{Appendix B: Computation of the excess hole excitation}
The excess hole excitation of the generated electron wavepackets can be determined by calculating the energy spectrum of each pulse. 
The electron-hole excitation spectrum, due to any periodic voltage pulse applied on the Fermi sea of a quantum conductor, can be computed in the framework of Floquet scattering theory \cite{Moskalets2002}. 
Under the effect of a voltage $V(t)$, the wavefunction acquires an extra phase term given by,
\begin{equation}
    \phi (t)  = \frac{e}{\hbar} \int^t_{-\infty} V(t')dt',
\end{equation}
where $e$ and $\hbar$ are the electron charge and the reduced Planck's constants, respectively. The $V(t)$ is the applied voltage pulse's temporal profile. In case of a periodic voltage pulse with repetition frequency $\nu$, the extra phase can be decomposed with the following Fourier series,
\begin{equation}
    \exp{\left(-i\phi (t)\right)}  = \sum_{l=-\infty}^{+\infty} p_l \exp{(-i2\pi l \nu t)}.
\end{equation}
where $|p_l|^2$ is the probability of one electron to absorb or emit $l$ photons. The number of electrons and holes can then be computed as \cite{Dubois2013a},
\begin{equation}
    N_e=\sum_{l=1}^{+\infty} l |p_l|^2,
\end{equation}
\begin{equation}
    N_h=\sum_{l=-\infty}^{-1} (-l) |p_l|^2.
\end{equation}\textbf{}

Figure \ref{suppfig:pulses_spectrum} shows the calculated electron-hole excitation spectrum of the generated voltage pulses of Figure~3~(a), (b), and~(d) of the main text. 
To accurately calculate the excitation spectrum, the amplitude of the measured voltage pulses has been numerically reduced and biased to contain one elementary charge per pulse. 
The excess hole excitation of each pulse, defined by the ratio of hole-to-electron ($N_{\rm{h}}/N_{\rm{e}}$) excitation, is very low and depends on the pulse duration.
For long-duration pulses, \SI{80}{\ps}, the ratio of the generated holes-to-electrons is minimal and in the \SI{0.07}{\percent} range. 
This corresponds to the generation of less than 1 hole for every $10^3$ electrons. 
Reducing the pulse duration to \SI{50}{\ps} increases slightly this ratio to \SI{0.22}{\percent}, whereas for a \SI{25}{\ps} pulse the ratio is \SI{0.55}{\percent}.
In Figure \ref{suppfig:pulses_spectrum} (f), we observe an increase on the amount of generated holes for large harmonics (i.e. $h\nu=-9$).
This effect is well-matched to our theory, and it is due to the lack of higher harmonics when synthesising short lorentzian voltage pulses. 
As explained in the main text, and shown in Figure \ref{suppfig:pulses_spectrum} (c), this also translates into satellite oscillations on either side of the main pulse. 

Analysing the energy spectrum of the generated electron wavepackets, using the Floquet scattering theory, provides quick information on the amount of holes contained within a wavepacket.
This approach, however, although it is faster than a tomography experiment, it cannot provide any additional information on whether the electrons contained within the wavepacket are in a pure or mixed quantum states \cite{Bisognin2019}.\\ \\ \\

\newpage

\nocite{Bisognin2019,Dubois2013a,Moskalets2002}
\bibliography{main}

\begin{thebibliography}{46}%
\makeatletter
\providecommand \@ifxundefined [1]{%
 \@ifx{#1\undefined}
}%
\providecommand \@ifnum [1]{%
 \ifnum #1\expandafter \@firstoftwo
 \else \expandafter \@secondoftwo
 \fi
}%
\providecommand \@ifx [1]{%
 \ifx #1\expandafter \@firstoftwo
 \else \expandafter \@secondoftwo
 \fi
}%
\providecommand \natexlab [1]{#1}%
\providecommand \enquote  [1]{``#1''}%
\providecommand \bibnamefont  [1]{#1}%
\providecommand \bibfnamefont [1]{#1}%
\providecommand \citenamefont [1]{#1}%
\providecommand \href@noop [0]{\@secondoftwo}%
\providecommand \href [0]{\begingroup \@sanitize@url \@href}%
\providecommand \@href[1]{\@@startlink{#1}\@@href}%
\providecommand \@@href[1]{\endgroup#1\@@endlink}%
\providecommand \@sanitize@url [0]{\catcode `\\12\catcode `\$12\catcode
  `\&12\catcode `\#12\catcode `\^12\catcode `\_12\catcode `\%12\relax}%
\providecommand \@@startlink[1]{}%
\providecommand \@@endlink[0]{}%
\providecommand \url  [0]{\begingroup\@sanitize@url \@url }%
\providecommand \@url [1]{\endgroup\@href {#1}{\urlprefix }}%
\providecommand \urlprefix  [0]{URL }%
\providecommand \Eprint [0]{\href }%
\providecommand \doibase [0]{https://doi.org/}%
\providecommand \selectlanguage [0]{\@gobble}%
\providecommand \bibinfo  [0]{\@secondoftwo}%
\providecommand \bibfield  [0]{\@secondoftwo}%
\providecommand \translation [1]{[#1]}%
\providecommand \BibitemOpen [0]{}%
\providecommand \bibitemStop [0]{}%
\providecommand \bibitemNoStop [0]{.\EOS\space}%
\providecommand \EOS [0]{\spacefactor3000\relax}%
\providecommand \BibitemShut  [1]{\csname bibitem#1\endcsname}%
\let\auto@bib@innerbib\@empty
\bibitem [{\citenamefont {Stern}(2008)}]{Stern2008}%
  \BibitemOpen
  \bibfield  {author} {\bibinfo {author} {\bibfnamefont {A.}~\bibnamefont
  {Stern}},\ }\bibfield  {title} {\bibinfo {title} {Anyons and the quantum
  {Hall} effect—{A} pedagogical review},\ }\href
  {https://doi.org/10.1016/j.aop.2007.10.008} {\bibfield  {journal} {\bibinfo
  {journal} {Annals of Physics}\ }\bibinfo {series} {January {Special} {Issue}
  2008},\ \textbf {\bibinfo {volume} {323}},\ \bibinfo {pages} {204} (\bibinfo
  {year} {2008})}\BibitemShut {NoStop}%
\bibitem [{\citenamefont {Oliver}\ \emph {et~al.}(1999)\citenamefont {Oliver},
  \citenamefont {Kim}, \citenamefont {Liu},\ and\ \citenamefont
  {Yamamoto}}]{Oliver1999}%
  \BibitemOpen
  \bibfield  {author} {\bibinfo {author} {\bibfnamefont {W.~D.}\ \bibnamefont
  {Oliver}}, \bibinfo {author} {\bibfnamefont {J.}~\bibnamefont {Kim}},
  \bibinfo {author} {\bibfnamefont {R.~C.}\ \bibnamefont {Liu}},\ and\ \bibinfo
  {author} {\bibfnamefont {Y.}~\bibnamefont {Yamamoto}},\ }\bibfield  {title}
  {\bibinfo {title} {Hanbury {Brown} and {Twiss}-{Type} {Experiment} with
  {Electrons}},\ }\href {https://doi.org/10.1126/science.284.5412.299}
  {\bibfield  {journal} {\bibinfo  {journal} {Science}\ }\textbf {\bibinfo
  {volume} {284}},\ \bibinfo {pages} {299} (\bibinfo {year}
  {1999})}\BibitemShut {NoStop}%
\bibitem [{\citenamefont {Henny}\ \emph {et~al.}(1999)\citenamefont {Henny},
  \citenamefont {Oberholzer}, \citenamefont {Strunk}, \citenamefont {Heinzel},
  \citenamefont {Ensslin}, \citenamefont {Holland},\ and\ \citenamefont
  {Sch\"onenberger}}]{Henny1999}%
  \BibitemOpen
  \bibfield  {author} {\bibinfo {author} {\bibfnamefont {M.}~\bibnamefont
  {Henny}}, \bibinfo {author} {\bibfnamefont {S.}~\bibnamefont {Oberholzer}},
  \bibinfo {author} {\bibfnamefont {C.}~\bibnamefont {Strunk}}, \bibinfo
  {author} {\bibfnamefont {T.}~\bibnamefont {Heinzel}}, \bibinfo {author}
  {\bibfnamefont {K.}~\bibnamefont {Ensslin}}, \bibinfo {author} {\bibfnamefont
  {M.}~\bibnamefont {Holland}},\ and\ \bibinfo {author} {\bibfnamefont
  {C.}~\bibnamefont {Sch\"onenberger}},\ }\bibfield  {title} {\bibinfo {title}
  {The {Fermionic} {Hanbury} {Brown} and {Twiss} {Experiment}},\ }\href
  {https://doi.org/10.1126/science.284.5412.296} {\bibfield  {journal}
  {\bibinfo  {journal} {Science}\ }\textbf {\bibinfo {volume} {284}},\ \bibinfo
  {pages} {296} (\bibinfo {year} {1999})}\BibitemShut {NoStop}%
\bibitem [{\citenamefont {Bocquillon}\ \emph {et~al.}(2013)\citenamefont
  {Bocquillon}, \citenamefont {Freulon}, \citenamefont {Berroir}, \citenamefont
  {Degiovanni}, \citenamefont {Pla{\c{c}}ais}, \citenamefont {Cavanna},
  \citenamefont {Jin},\ and\ \citenamefont {F{\`{e}}ve}}]{Bocquillon2013}%
  \BibitemOpen
  \bibfield  {author} {\bibinfo {author} {\bibfnamefont {E.}~\bibnamefont
  {Bocquillon}}, \bibinfo {author} {\bibfnamefont {V.}~\bibnamefont {Freulon}},
  \bibinfo {author} {\bibfnamefont {J.-M.}\ \bibnamefont {Berroir}}, \bibinfo
  {author} {\bibfnamefont {P.}~\bibnamefont {Degiovanni}}, \bibinfo {author}
  {\bibfnamefont {B.}~\bibnamefont {Pla{\c{c}}ais}}, \bibinfo {author}
  {\bibfnamefont {A.}~\bibnamefont {Cavanna}}, \bibinfo {author} {\bibfnamefont
  {Y.}~\bibnamefont {Jin}},\ and\ \bibinfo {author} {\bibfnamefont
  {G.}~\bibnamefont {F{\`{e}}ve}},\ }\bibfield  {title} {\bibinfo {title}
  {Coherence and indistinguishability of single electrons emitted by
  independent sources},\ }\href {https://doi.org/10.1126/science.1232572}
  {\bibfield  {journal} {\bibinfo  {journal} {Science}\ }\textbf {\bibinfo
  {volume} {339}},\ \bibinfo {pages} {1054} (\bibinfo {year}
  {2013})}\BibitemShut {NoStop}%
\bibitem [{\citenamefont {Dubois}\ \emph
  {et~al.}(2013{\natexlab{a}})\citenamefont {Dubois}, \citenamefont {Jullien},
  \citenamefont {Portier}, \citenamefont {Roche}, \citenamefont {Cavanna},
  \citenamefont {Jin}, \citenamefont {Wegscheider}, \citenamefont {Roulleau},\
  and\ \citenamefont {Glattli}}]{Dubois2013}%
  \BibitemOpen
  \bibfield  {author} {\bibinfo {author} {\bibfnamefont {J.}~\bibnamefont
  {Dubois}}, \bibinfo {author} {\bibfnamefont {T.}~\bibnamefont {Jullien}},
  \bibinfo {author} {\bibfnamefont {F.}~\bibnamefont {Portier}}, \bibinfo
  {author} {\bibfnamefont {P.}~\bibnamefont {Roche}}, \bibinfo {author}
  {\bibfnamefont {A.}~\bibnamefont {Cavanna}}, \bibinfo {author} {\bibfnamefont
  {Y.}~\bibnamefont {Jin}}, \bibinfo {author} {\bibfnamefont {W.}~\bibnamefont
  {Wegscheider}}, \bibinfo {author} {\bibfnamefont {P.}~\bibnamefont
  {Roulleau}},\ and\ \bibinfo {author} {\bibfnamefont {D.~C.}\ \bibnamefont
  {Glattli}},\ }\bibfield  {title} {\bibinfo {title} {Minimal-excitation states
  for electron quantum optics using levitons},\ }\href
  {https://doi.org/10.1038/nature12713} {\bibfield  {journal} {\bibinfo
  {journal} {Nature}\ }\textbf {\bibinfo {volume} {502}},\ \bibinfo {pages}
  {659} (\bibinfo {year} {2013}{\natexlab{a}})}\BibitemShut {NoStop}%
\bibitem [{\citenamefont {Freulon}\ \emph {et~al.}(2015)\citenamefont
  {Freulon}, \citenamefont {Marguerite}, \citenamefont {Berroir}, \citenamefont
  {Plaçais}, \citenamefont {Cavanna}, \citenamefont {Jon},\ and\ \citenamefont
  {Fève}}]{Freulon2015}%
  \BibitemOpen
  \bibfield  {author} {\bibinfo {author} {\bibfnamefont {V.}~\bibnamefont
  {Freulon}}, \bibinfo {author} {\bibfnamefont {A.}~\bibnamefont {Marguerite}},
  \bibinfo {author} {\bibfnamefont {J.-M.}\ \bibnamefont {Berroir}}, \bibinfo
  {author} {\bibfnamefont {B.}~\bibnamefont {Plaçais}}, \bibinfo {author}
  {\bibfnamefont {A.}~\bibnamefont {Cavanna}}, \bibinfo {author} {\bibfnamefont
  {Y.}~\bibnamefont {Jon}},\ and\ \bibinfo {author} {\bibfnamefont
  {G.}~\bibnamefont {Fève}},\ }\bibfield  {title} {\bibinfo {title}
  {Hong-{Ou}-{Mandel} experiment for temporal investigation of single-electron
  fractionalization},\ }\href {https://doi.org/10.1038/ncomms7854} {\bibfield
  {journal} {\bibinfo  {journal} {Nature Communications}\ }\textbf {\bibinfo
  {volume} {6}},\ \bibinfo {pages} {6854} (\bibinfo {year} {2015})}\BibitemShut
  {NoStop}%
\bibitem [{\citenamefont {Kapfer}\ \emph {et~al.}(2019)\citenamefont {Kapfer},
  \citenamefont {Roulleau}, \citenamefont {Santin}, \citenamefont {Farrer},
  \citenamefont {Ritchie},\ and\ \citenamefont {Glattli}}]{Kapfer2019}%
  \BibitemOpen
  \bibfield  {author} {\bibinfo {author} {\bibfnamefont {M.}~\bibnamefont
  {Kapfer}}, \bibinfo {author} {\bibfnamefont {P.}~\bibnamefont {Roulleau}},
  \bibinfo {author} {\bibfnamefont {M.}~\bibnamefont {Santin}}, \bibinfo
  {author} {\bibfnamefont {I.}~\bibnamefont {Farrer}}, \bibinfo {author}
  {\bibfnamefont {D.~A.}\ \bibnamefont {Ritchie}},\ and\ \bibinfo {author}
  {\bibfnamefont {D.~C.}\ \bibnamefont {Glattli}},\ }\bibfield  {title}
  {\bibinfo {title} {A {Josephson} relation for fractionally charged anyons},\
  }\href {https://doi.org/10.1126/science.aau3539} {\bibfield  {journal}
  {\bibinfo  {journal} {Science}\ }\textbf {\bibinfo {volume} {363}},\ \bibinfo
  {pages} {846} (\bibinfo {year} {2019})}\BibitemShut {NoStop}%
\bibitem [{\citenamefont {Bartolomei}\ \emph {et~al.}(2020)\citenamefont
  {Bartolomei}, \citenamefont {Kumar}, \citenamefont {Bisognin}, \citenamefont
  {Marguerite}, \citenamefont {Berroir}, \citenamefont {Bocquillon},
  \citenamefont {Plaçais}, \citenamefont {Cavanna}, \citenamefont {Dong},
  \citenamefont {Gennser}, \citenamefont {Jin},\ and\ \citenamefont
  {Fève}}]{Bartolomei2020}%
  \BibitemOpen
  \bibfield  {author} {\bibinfo {author} {\bibfnamefont {H.}~\bibnamefont
  {Bartolomei}}, \bibinfo {author} {\bibfnamefont {M.}~\bibnamefont {Kumar}},
  \bibinfo {author} {\bibfnamefont {R.}~\bibnamefont {Bisognin}}, \bibinfo
  {author} {\bibfnamefont {A.}~\bibnamefont {Marguerite}}, \bibinfo {author}
  {\bibfnamefont {J.-M.}\ \bibnamefont {Berroir}}, \bibinfo {author}
  {\bibfnamefont {E.}~\bibnamefont {Bocquillon}}, \bibinfo {author}
  {\bibfnamefont {B.}~\bibnamefont {Plaçais}}, \bibinfo {author}
  {\bibfnamefont {A.}~\bibnamefont {Cavanna}}, \bibinfo {author} {\bibfnamefont
  {Q.}~\bibnamefont {Dong}}, \bibinfo {author} {\bibfnamefont {U.}~\bibnamefont
  {Gennser}}, \bibinfo {author} {\bibfnamefont {Y.}~\bibnamefont {Jin}},\ and\
  \bibinfo {author} {\bibfnamefont {G.}~\bibnamefont {Fève}},\ }\bibfield
  {title} {\bibinfo {title} {Fractional statistics in anyon collisions},\
  }\href {https://doi.org/10.1126/science.aaz5601} {\bibfield  {journal}
  {\bibinfo  {journal} {Science}\ }\textbf {\bibinfo {volume} {368}},\ \bibinfo
  {pages} {173} (\bibinfo {year} {2020})}\BibitemShut {NoStop}%
\bibitem [{\citenamefont {Nakamura}\ \emph {et~al.}(2020)\citenamefont
  {Nakamura}, \citenamefont {Liang}, \citenamefont {Gardner},\ and\
  \citenamefont {Manfra}}]{Nakamura2020}%
  \BibitemOpen
  \bibfield  {author} {\bibinfo {author} {\bibfnamefont {J.}~\bibnamefont
  {Nakamura}}, \bibinfo {author} {\bibfnamefont {S.}~\bibnamefont {Liang}},
  \bibinfo {author} {\bibfnamefont {G.~C.}\ \bibnamefont {Gardner}},\ and\
  \bibinfo {author} {\bibfnamefont {M.~J.}\ \bibnamefont {Manfra}},\ }\bibfield
   {title} {\bibinfo {title} {Direct observation of anyonic braiding
  statistics},\ }\href {https://doi.org/10.1038/s41567-020-1019-1} {\bibfield
  {journal} {\bibinfo  {journal} {Nature Physics}\ }\textbf {\bibinfo {volume}
  {16}},\ \bibinfo {pages} {931} (\bibinfo {year} {2020})}\BibitemShut
  {NoStop}%
\bibitem [{\citenamefont {Taktak}\ \emph {et~al.}(2022)\citenamefont {Taktak},
  \citenamefont {Kapfer}, \citenamefont {Nath}, \citenamefont {Roulleau},
  \citenamefont {Acciai}, \citenamefont {Splettstoesser}, \citenamefont
  {Farrer}, \citenamefont {Ritchie},\ and\ \citenamefont
  {Glattli}}]{Taktak2022}%
  \BibitemOpen
  \bibfield  {author} {\bibinfo {author} {\bibfnamefont {I.}~\bibnamefont
  {Taktak}}, \bibinfo {author} {\bibfnamefont {M.}~\bibnamefont {Kapfer}},
  \bibinfo {author} {\bibfnamefont {J.}~\bibnamefont {Nath}}, \bibinfo {author}
  {\bibfnamefont {P.}~\bibnamefont {Roulleau}}, \bibinfo {author}
  {\bibfnamefont {M.}~\bibnamefont {Acciai}}, \bibinfo {author} {\bibfnamefont
  {J.}~\bibnamefont {Splettstoesser}}, \bibinfo {author} {\bibfnamefont
  {I.}~\bibnamefont {Farrer}}, \bibinfo {author} {\bibfnamefont {D.~A.}\
  \bibnamefont {Ritchie}},\ and\ \bibinfo {author} {\bibfnamefont {D.~C.}\
  \bibnamefont {Glattli}},\ }\bibfield  {title} {\bibinfo {title} {Two-particle
  time-domain interferometry in the fractional quantum {Hall} effect regime},\
  }\href {https://doi.org/10.1038/s41467-022-33603-3} {\bibfield  {journal}
  {\bibinfo  {journal} {Nature Communications}\ }\textbf {\bibinfo {volume}
  {13}},\ \bibinfo {pages} {5863} (\bibinfo {year} {2022})}\BibitemShut
  {NoStop}%
\bibitem [{\citenamefont {Kamata}\ \emph {et~al.}(2022)\citenamefont {Kamata},
  \citenamefont {Irie}, \citenamefont {Kumada},\ and\ \citenamefont
  {Muraki}}]{Kamata2022}%
  \BibitemOpen
  \bibfield  {author} {\bibinfo {author} {\bibfnamefont {H.}~\bibnamefont
  {Kamata}}, \bibinfo {author} {\bibfnamefont {H.}~\bibnamefont {Irie}},
  \bibinfo {author} {\bibfnamefont {N.}~\bibnamefont {Kumada}},\ and\ \bibinfo
  {author} {\bibfnamefont {K.}~\bibnamefont {Muraki}},\ }\bibfield  {title}
  {\bibinfo {title} {Time-resolved measurement of ambipolar edge magnetoplasmon
  transport in {InAs}/{InGaSb} composite quantum wells},\ }\href
  {https://doi.org/10.1103/PhysRevResearch.4.033214} {\bibfield  {journal}
  {\bibinfo  {journal} {Physical Review Research}\ }\textbf {\bibinfo {volume}
  {4}},\ \bibinfo {pages} {033214} (\bibinfo {year} {2022})}\BibitemShut
  {NoStop}%
\bibitem [{\citenamefont {Giblin}\ \emph {et~al.}(2020)\citenamefont {Giblin},
  \citenamefont {Mykkänen}, \citenamefont {Kemppinen}, \citenamefont
  {Immonen}, \citenamefont {Manninen}, \citenamefont {Jenei}, \citenamefont
  {Möttönen}, \citenamefont {Yamahata}, \citenamefont {Fujiwara},\ and\
  \citenamefont {Kataoka}}]{Giblin2020}%
  \BibitemOpen
  \bibfield  {author} {\bibinfo {author} {\bibfnamefont {S.~P.}\ \bibnamefont
  {Giblin}}, \bibinfo {author} {\bibfnamefont {E.}~\bibnamefont {Mykkänen}},
  \bibinfo {author} {\bibfnamefont {A.}~\bibnamefont {Kemppinen}}, \bibinfo
  {author} {\bibfnamefont {P.}~\bibnamefont {Immonen}}, \bibinfo {author}
  {\bibfnamefont {A.}~\bibnamefont {Manninen}}, \bibinfo {author}
  {\bibfnamefont {M.}~\bibnamefont {Jenei}}, \bibinfo {author} {\bibfnamefont
  {M.}~\bibnamefont {Möttönen}}, \bibinfo {author} {\bibfnamefont
  {G.}~\bibnamefont {Yamahata}}, \bibinfo {author} {\bibfnamefont
  {A.}~\bibnamefont {Fujiwara}},\ and\ \bibinfo {author} {\bibfnamefont
  {M.}~\bibnamefont {Kataoka}},\ }\bibfield  {title} {\bibinfo {title}
  {Realisation of a quantum current standard at liquid helium temperature with
  sub-ppm reproducibility},\ }\href {https://doi.org/10.1088/1681-7575/ab72e0}
  {\bibfield  {journal} {\bibinfo  {journal} {Metrologia}\ }\textbf {\bibinfo
  {volume} {57}},\ \bibinfo {pages} {025013} (\bibinfo {year}
  {2020})}\BibitemShut {NoStop}%
\bibitem [{\citenamefont {Scherer}\ and\ \citenamefont
  {Schumacher}(2019)}]{Scherer2019}%
  \BibitemOpen
  \bibfield  {author} {\bibinfo {author} {\bibfnamefont {H.}~\bibnamefont
  {Scherer}}\ and\ \bibinfo {author} {\bibfnamefont {H.~W.}\ \bibnamefont
  {Schumacher}},\ }\bibfield  {title} {\bibinfo {title} {Single-electron pumps
  and quantum current metrology in~the~revised {SI}},\ }\href
  {https://doi.org/10.1002/andp.201800371} {\bibfield  {journal} {\bibinfo
  {journal} {Annalen der Physik}\ }\textbf {\bibinfo {volume} {531}},\ \bibinfo
  {pages} {1800371} (\bibinfo {year} {2019})}\BibitemShut {NoStop}%
\bibitem [{\citenamefont {Kaneko}\ \emph {et~al.}(2016)\citenamefont {Kaneko},
  \citenamefont {Nakamura},\ and\ \citenamefont {Okazaki}}]{Kaneko2016}%
  \BibitemOpen
  \bibfield  {author} {\bibinfo {author} {\bibfnamefont {N.-H.}\ \bibnamefont
  {Kaneko}}, \bibinfo {author} {\bibfnamefont {S.}~\bibnamefont {Nakamura}},\
  and\ \bibinfo {author} {\bibfnamefont {Y.}~\bibnamefont {Okazaki}},\
  }\bibfield  {title} {\bibinfo {title} {A review of the quantum current
  standard},\ }\href {https://doi.org/10.1088/0957-0233/27/3/032001} {\bibfield
   {journal} {\bibinfo  {journal} {Measurement Science and Technology}\
  }\textbf {\bibinfo {volume} {27}},\ \bibinfo {pages} {032001} (\bibinfo
  {year} {2016})}\BibitemShut {NoStop}%
\bibitem [{\citenamefont {Bäuerle}\ \emph {et~al.}(2018)\citenamefont
  {Bäuerle}, \citenamefont {Glattli}, \citenamefont {Meunier}, \citenamefont
  {Portier}, \citenamefont {Roche}, \citenamefont {Roulleau}, \citenamefont
  {Takada},\ and\ \citenamefont {Waintal}}]{Baeuerle2018}%
  \BibitemOpen
  \bibfield  {author} {\bibinfo {author} {\bibfnamefont {C.}~\bibnamefont
  {Bäuerle}}, \bibinfo {author} {\bibfnamefont {D.~C.}\ \bibnamefont
  {Glattli}}, \bibinfo {author} {\bibfnamefont {T.}~\bibnamefont {Meunier}},
  \bibinfo {author} {\bibfnamefont {F.}~\bibnamefont {Portier}}, \bibinfo
  {author} {\bibfnamefont {P.}~\bibnamefont {Roche}}, \bibinfo {author}
  {\bibfnamefont {P.}~\bibnamefont {Roulleau}}, \bibinfo {author}
  {\bibfnamefont {S.}~\bibnamefont {Takada}},\ and\ \bibinfo {author}
  {\bibfnamefont {X.}~\bibnamefont {Waintal}},\ }\bibfield  {title} {\bibinfo
  {title} {Coherent control of single electrons: a review of current
  progress},\ }\href@noop {} {\bibfield  {journal} {\bibinfo  {journal}
  {Reports on Progress in Physics}\ }\textbf {\bibinfo {volume} {81}},\
  \bibinfo {pages} {056503} (\bibinfo {year} {2018})}\BibitemShut {NoStop}%
\bibitem [{\citenamefont {Edlbauer}\ \emph {et~al.}(2022)\citenamefont
  {Edlbauer}, \citenamefont {Wang}, \citenamefont {Crozes}, \citenamefont
  {Perrier}, \citenamefont {Ouacel}, \citenamefont {Geffroy}, \citenamefont
  {Georgiou}, \citenamefont {Chatzikyriakou}, \citenamefont {Lacerda-Santos},
  \citenamefont {Waintal}, \citenamefont {Glattli}, \citenamefont {Roulleau},
  \citenamefont {Nath}, \citenamefont {Kataoka}, \citenamefont
  {Splettstoesser}, \citenamefont {Acciai}, \citenamefont {da~Silva~Figueira},
  \citenamefont {Öztas}, \citenamefont {Trellakis}, \citenamefont {Grange},
  \citenamefont {Yevtushenko}, \citenamefont {Birner},\ and\ \citenamefont
  {Bäuerle}}]{Edlbauer2022}%
  \BibitemOpen
  \bibfield  {author} {\bibinfo {author} {\bibfnamefont {H.}~\bibnamefont
  {Edlbauer}}, \bibinfo {author} {\bibfnamefont {J.}~\bibnamefont {Wang}},
  \bibinfo {author} {\bibfnamefont {T.}~\bibnamefont {Crozes}}, \bibinfo
  {author} {\bibfnamefont {P.}~\bibnamefont {Perrier}}, \bibinfo {author}
  {\bibfnamefont {S.}~\bibnamefont {Ouacel}}, \bibinfo {author} {\bibfnamefont
  {C.}~\bibnamefont {Geffroy}}, \bibinfo {author} {\bibfnamefont
  {G.}~\bibnamefont {Georgiou}}, \bibinfo {author} {\bibfnamefont
  {E.}~\bibnamefont {Chatzikyriakou}}, \bibinfo {author} {\bibfnamefont
  {A.}~\bibnamefont {Lacerda-Santos}}, \bibinfo {author} {\bibfnamefont
  {X.}~\bibnamefont {Waintal}}, \bibinfo {author} {\bibfnamefont {D.~C.}\
  \bibnamefont {Glattli}}, \bibinfo {author} {\bibfnamefont {P.}~\bibnamefont
  {Roulleau}}, \bibinfo {author} {\bibfnamefont {J.}~\bibnamefont {Nath}},
  \bibinfo {author} {\bibfnamefont {M.}~\bibnamefont {Kataoka}}, \bibinfo
  {author} {\bibfnamefont {J.}~\bibnamefont {Splettstoesser}}, \bibinfo
  {author} {\bibfnamefont {M.}~\bibnamefont {Acciai}}, \bibinfo {author}
  {\bibfnamefont {M.~C.}\ \bibnamefont {da~Silva~Figueira}}, \bibinfo {author}
  {\bibfnamefont {K.}~\bibnamefont {Öztas}}, \bibinfo {author} {\bibfnamefont
  {A.}~\bibnamefont {Trellakis}}, \bibinfo {author} {\bibfnamefont
  {T.}~\bibnamefont {Grange}}, \bibinfo {author} {\bibfnamefont {O.~M.}\
  \bibnamefont {Yevtushenko}}, \bibinfo {author} {\bibfnamefont
  {S.}~\bibnamefont {Birner}},\ and\ \bibinfo {author} {\bibfnamefont
  {C.}~\bibnamefont {Bäuerle}},\ }\bibfield  {title} {\bibinfo {title}
  {Semiconductor-based electron flying qubits: review on recent progress
  accelerated by numerical modelling},\ }\bibfield  {journal} {\bibinfo
  {journal} {{EPJ} Quantum Technology}\ }\textbf {\bibinfo {volume} {9}},\
  \href@noop {} {} (\bibinfo {year} {2022})\BibitemShut {NoStop}%
\bibitem [{\citenamefont {Takada}\ \emph {et~al.}(2019)\citenamefont {Takada},
  \citenamefont {Edlbauer}, \citenamefont {Lepage}, \citenamefont {Wang},
  \citenamefont {Mortemousque}, \citenamefont {Georgiou}, \citenamefont
  {Barnes}, \citenamefont {Ford}, \citenamefont {Yuan}, \citenamefont {Santos},
  \citenamefont {Waintal}, \citenamefont {Ludwig}, \citenamefont {Wieck},
  \citenamefont {Urdampilleta}, \citenamefont {Meunier},\ and\ \citenamefont
  {Bäuerle}}]{Takada2019}%
  \BibitemOpen
  \bibfield  {author} {\bibinfo {author} {\bibfnamefont {S.}~\bibnamefont
  {Takada}}, \bibinfo {author} {\bibfnamefont {H.}~\bibnamefont {Edlbauer}},
  \bibinfo {author} {\bibfnamefont {H.~V.}\ \bibnamefont {Lepage}}, \bibinfo
  {author} {\bibfnamefont {J.}~\bibnamefont {Wang}}, \bibinfo {author}
  {\bibfnamefont {P.-A.}\ \bibnamefont {Mortemousque}}, \bibinfo {author}
  {\bibfnamefont {G.}~\bibnamefont {Georgiou}}, \bibinfo {author}
  {\bibfnamefont {C.~H.~W.}\ \bibnamefont {Barnes}}, \bibinfo {author}
  {\bibfnamefont {C.~J.~B.}\ \bibnamefont {Ford}}, \bibinfo {author}
  {\bibfnamefont {M.}~\bibnamefont {Yuan}}, \bibinfo {author} {\bibfnamefont
  {P.~V.}\ \bibnamefont {Santos}}, \bibinfo {author} {\bibfnamefont
  {X.}~\bibnamefont {Waintal}}, \bibinfo {author} {\bibfnamefont
  {A.}~\bibnamefont {Ludwig}}, \bibinfo {author} {\bibfnamefont {A.~D.}\
  \bibnamefont {Wieck}}, \bibinfo {author} {\bibfnamefont {M.}~\bibnamefont
  {Urdampilleta}}, \bibinfo {author} {\bibfnamefont {T.}~\bibnamefont
  {Meunier}},\ and\ \bibinfo {author} {\bibfnamefont {C.}~\bibnamefont
  {Bäuerle}},\ }\bibfield  {title} {\bibinfo {title} {Sound-driven
  single-electron transfer in a circuit of coupled quantum rails},\ }\href
  {https://doi.org/10.1038/s41467-019-12514-w} {\bibfield  {journal} {\bibinfo
  {journal} {Nature Communications}\ }\textbf {\bibinfo {volume} {10}},\
  \bibinfo {pages} {4557} (\bibinfo {year} {2019})}\BibitemShut {NoStop}%
\bibitem [{\citenamefont {Wang}\ \emph {et~al.}(2022)\citenamefont {Wang},
  \citenamefont {Ota}, \citenamefont {Edlbauer}, \citenamefont {Jadot},
  \citenamefont {Mortemousque}, \citenamefont {Richard}, \citenamefont
  {Okazaki}, \citenamefont {Nakamura}, \citenamefont {Ludwig}, \citenamefont
  {Wieck}, \citenamefont {Urdampilleta}, \citenamefont {Meunier}, \citenamefont
  {Kodera}, \citenamefont {Kaneko}, \citenamefont {Takada},\ and\ \citenamefont
  {Bäuerle}}]{Wang2022}%
  \BibitemOpen
  \bibfield  {author} {\bibinfo {author} {\bibfnamefont {J.}~\bibnamefont
  {Wang}}, \bibinfo {author} {\bibfnamefont {S.}~\bibnamefont {Ota}}, \bibinfo
  {author} {\bibfnamefont {H.}~\bibnamefont {Edlbauer}}, \bibinfo {author}
  {\bibfnamefont {B.}~\bibnamefont {Jadot}}, \bibinfo {author} {\bibfnamefont
  {P.-A.}\ \bibnamefont {Mortemousque}}, \bibinfo {author} {\bibfnamefont
  {A.}~\bibnamefont {Richard}}, \bibinfo {author} {\bibfnamefont
  {Y.}~\bibnamefont {Okazaki}}, \bibinfo {author} {\bibfnamefont
  {S.}~\bibnamefont {Nakamura}}, \bibinfo {author} {\bibfnamefont
  {A.}~\bibnamefont {Ludwig}}, \bibinfo {author} {\bibfnamefont {A.~D.}\
  \bibnamefont {Wieck}}, \bibinfo {author} {\bibfnamefont {M.}~\bibnamefont
  {Urdampilleta}}, \bibinfo {author} {\bibfnamefont {T.}~\bibnamefont
  {Meunier}}, \bibinfo {author} {\bibfnamefont {T.}~\bibnamefont {Kodera}},
  \bibinfo {author} {\bibfnamefont {N.-H.}\ \bibnamefont {Kaneko}}, \bibinfo
  {author} {\bibfnamefont {S.}~\bibnamefont {Takada}},\ and\ \bibinfo {author}
  {\bibfnamefont {C.}~\bibnamefont {Bäuerle}},\ }\bibfield  {title} {\bibinfo
  {title} {Generation of a {Single}-{Cycle} {Acoustic} {Pulse}: {A} {Scalable}
  {Solution} for {Transport} in {Single}-{Electron} {Circuits}},\ }\href
  {https://doi.org/10.1103/PhysRevX.12.031035} {\bibfield  {journal} {\bibinfo
  {journal} {Physical Review X}\ }\textbf {\bibinfo {volume} {12}},\ \bibinfo
  {pages} {031035} (\bibinfo {year} {2022})}\BibitemShut {NoStop}%
\bibitem [{\citenamefont {Wang}\ \emph {et~al.}()\citenamefont {Wang},
  \citenamefont {Edlbauer}, \citenamefont {Richard}, \citenamefont {Ota},
  \citenamefont {Park}, \citenamefont {Shim}, \citenamefont {Ludwig},
  \citenamefont {Wieck}, \citenamefont {Sim}, \citenamefont {Urdampilleta},
  \citenamefont {Meunier}, \citenamefont {Kodera}, \citenamefont {Kaneko},
  \citenamefont {Sellier}, \citenamefont {Waintal}, \citenamefont {Takada},\
  and\ \citenamefont {Bäuerle}}]{Wang2022a}%
  \BibitemOpen
  \bibfield  {author} {\bibinfo {author} {\bibfnamefont {J.}~\bibnamefont
  {Wang}}, \bibinfo {author} {\bibfnamefont {H.}~\bibnamefont {Edlbauer}},
  \bibinfo {author} {\bibfnamefont {A.}~\bibnamefont {Richard}}, \bibinfo
  {author} {\bibfnamefont {S.}~\bibnamefont {Ota}}, \bibinfo {author}
  {\bibfnamefont {W.}~\bibnamefont {Park}}, \bibinfo {author} {\bibfnamefont
  {J.}~\bibnamefont {Shim}}, \bibinfo {author} {\bibfnamefont {A.}~\bibnamefont
  {Ludwig}}, \bibinfo {author} {\bibfnamefont {A.}~\bibnamefont {Wieck}},
  \bibinfo {author} {\bibfnamefont {H.-S.}\ \bibnamefont {Sim}}, \bibinfo
  {author} {\bibfnamefont {M.}~\bibnamefont {Urdampilleta}}, \bibinfo {author}
  {\bibfnamefont {T.}~\bibnamefont {Meunier}}, \bibinfo {author} {\bibfnamefont
  {T.}~\bibnamefont {Kodera}}, \bibinfo {author} {\bibfnamefont {N.-H.}\
  \bibnamefont {Kaneko}}, \bibinfo {author} {\bibfnamefont {H.}~\bibnamefont
  {Sellier}}, \bibinfo {author} {\bibfnamefont {X.}~\bibnamefont {Waintal}},
  \bibinfo {author} {\bibfnamefont {S.}~\bibnamefont {Takada}},\ and\ \bibinfo
  {author} {\bibfnamefont {C.}~\bibnamefont {Bäuerle}},\ }\href
  {https://doi.org/10.48550/arXiv.2210.03452} {\bibinfo {title}
  {Coulomb-mediated antibunching of an electron pair surfing on sound}},\
  \bibinfo {note} {arXiv:2210.03452 (2022)}\BibitemShut {NoStop}%
\bibitem [{\citenamefont {Ubbelohde}\ \emph {et~al.}()\citenamefont
  {Ubbelohde}, \citenamefont {Freise}, \citenamefont {Pavlovska}, \citenamefont
  {Silvestrov}, \citenamefont {Recher}, \citenamefont {Kokainis}, \citenamefont
  {Barinovs}, \citenamefont {Hohls}, \citenamefont {Weimann}, \citenamefont
  {Pierz},\ and\ \citenamefont {Kashcheyevs}}]{Ubbelohde2022}%
  \BibitemOpen
  \bibfield  {author} {\bibinfo {author} {\bibfnamefont {N.}~\bibnamefont
  {Ubbelohde}}, \bibinfo {author} {\bibfnamefont {L.}~\bibnamefont {Freise}},
  \bibinfo {author} {\bibfnamefont {E.}~\bibnamefont {Pavlovska}}, \bibinfo
  {author} {\bibfnamefont {P.~G.}\ \bibnamefont {Silvestrov}}, \bibinfo
  {author} {\bibfnamefont {P.}~\bibnamefont {Recher}}, \bibinfo {author}
  {\bibfnamefont {M.}~\bibnamefont {Kokainis}}, \bibinfo {author}
  {\bibfnamefont {G.}~\bibnamefont {Barinovs}}, \bibinfo {author}
  {\bibfnamefont {F.}~\bibnamefont {Hohls}}, \bibinfo {author} {\bibfnamefont
  {T.}~\bibnamefont {Weimann}}, \bibinfo {author} {\bibfnamefont
  {K.}~\bibnamefont {Pierz}},\ and\ \bibinfo {author} {\bibfnamefont
  {V.}~\bibnamefont {Kashcheyevs}},\ }\href
  {https://doi.org/10.48550/arXiv.2210.03632} {\bibinfo {title} {Two electrons
  interacting at a mesoscopic beam splitter}},\ \bibinfo {note}
  {arXiv:2210.03632 (2022)}\BibitemShut {NoStop}%
\bibitem [{\citenamefont {Fletcher}\ \emph {et~al.}()\citenamefont {Fletcher},
  \citenamefont {Park}, \citenamefont {Ryu}, \citenamefont {See}, \citenamefont
  {Griffiths}, \citenamefont {Jones}, \citenamefont {Farrer}, \citenamefont
  {Ritchie}, \citenamefont {Sim},\ and\ \citenamefont
  {Kataoka}}]{Fletcher2022}%
  \BibitemOpen
  \bibfield  {author} {\bibinfo {author} {\bibfnamefont {J.~D.}\ \bibnamefont
  {Fletcher}}, \bibinfo {author} {\bibfnamefont {W.}~\bibnamefont {Park}},
  \bibinfo {author} {\bibfnamefont {S.}~\bibnamefont {Ryu}}, \bibinfo {author}
  {\bibfnamefont {P.}~\bibnamefont {See}}, \bibinfo {author} {\bibfnamefont
  {J.~P.}\ \bibnamefont {Griffiths}}, \bibinfo {author} {\bibfnamefont
  {G.~A.~C.}\ \bibnamefont {Jones}}, \bibinfo {author} {\bibfnamefont
  {I.}~\bibnamefont {Farrer}}, \bibinfo {author} {\bibfnamefont {D.~A.}\
  \bibnamefont {Ritchie}}, \bibinfo {author} {\bibfnamefont {H.-S.}\
  \bibnamefont {Sim}},\ and\ \bibinfo {author} {\bibfnamefont {M.}~\bibnamefont
  {Kataoka}},\ }\href {https://doi.org/10.48550/arXiv.2210.03473} {\bibinfo
  {title} {Time-resolved {Coulomb} collision of single electrons}},\ \bibinfo
  {note} {arXiv:2210.03473 (2022)}\BibitemShut {NoStop}%
\bibitem [{\citenamefont {Jadot}\ \emph {et~al.}(2021)\citenamefont {Jadot},
  \citenamefont {Mortemousque}, \citenamefont {Chanrion}, \citenamefont
  {Thiney}, \citenamefont {Ludwig}, \citenamefont {Wieck}, \citenamefont
  {Urdampilleta}, \citenamefont {Bäuerle},\ and\ \citenamefont
  {Meunier}}]{Jadot2021}%
  \BibitemOpen
  \bibfield  {author} {\bibinfo {author} {\bibfnamefont {B.}~\bibnamefont
  {Jadot}}, \bibinfo {author} {\bibfnamefont {P.-A.}\ \bibnamefont
  {Mortemousque}}, \bibinfo {author} {\bibfnamefont {E.}~\bibnamefont
  {Chanrion}}, \bibinfo {author} {\bibfnamefont {V.}~\bibnamefont {Thiney}},
  \bibinfo {author} {\bibfnamefont {A.}~\bibnamefont {Ludwig}}, \bibinfo
  {author} {\bibfnamefont {A.~D.}\ \bibnamefont {Wieck}}, \bibinfo {author}
  {\bibfnamefont {M.}~\bibnamefont {Urdampilleta}}, \bibinfo {author}
  {\bibfnamefont {C.}~\bibnamefont {Bäuerle}},\ and\ \bibinfo {author}
  {\bibfnamefont {T.}~\bibnamefont {Meunier}},\ }\bibfield  {title} {\bibinfo
  {title} {Distant spin entanglement via fast and coherent electron
  shuttling},\ }\href {https://doi.org/10.1038/s41565-021-00846-y} {\bibfield
  {journal} {\bibinfo  {journal} {Nature Nanotechnology}\ }\textbf {\bibinfo
  {volume} {16}},\ \bibinfo {pages} {570} (\bibinfo {year} {2021})}\BibitemShut
  {NoStop}%
\bibitem [{\citenamefont {Fève}\ \emph {et~al.}(2007)\citenamefont {Fève},
  \citenamefont {Mahé}, \citenamefont {Berroir}, \citenamefont {Kontos},
  \citenamefont {Plaçais}, \citenamefont {Glattli}, \citenamefont {Cavanna},
  \citenamefont {Etienne},\ and\ \citenamefont {Jin}}]{Feve2007}%
  \BibitemOpen
  \bibfield  {author} {\bibinfo {author} {\bibfnamefont {G.}~\bibnamefont
  {Fève}}, \bibinfo {author} {\bibfnamefont {A.}~\bibnamefont {Mahé}},
  \bibinfo {author} {\bibfnamefont {J.-M.}\ \bibnamefont {Berroir}}, \bibinfo
  {author} {\bibfnamefont {T.}~\bibnamefont {Kontos}}, \bibinfo {author}
  {\bibfnamefont {B.}~\bibnamefont {Plaçais}}, \bibinfo {author}
  {\bibfnamefont {D.~C.}\ \bibnamefont {Glattli}}, \bibinfo {author}
  {\bibfnamefont {A.}~\bibnamefont {Cavanna}}, \bibinfo {author} {\bibfnamefont
  {B.}~\bibnamefont {Etienne}},\ and\ \bibinfo {author} {\bibfnamefont
  {Y.}~\bibnamefont {Jin}},\ }\bibfield  {title} {\bibinfo {title} {An
  on-demand coherent single-electron source},\ }\href
  {https://doi.org/10.1126/science.1141243} {\bibfield  {journal} {\bibinfo
  {journal} {Science}\ }\textbf {\bibinfo {volume} {316}},\ \bibinfo {pages}
  {1169} (\bibinfo {year} {2007})}\BibitemShut {NoStop}%
\bibitem [{\citenamefont {Blumenthal}\ \emph {et~al.}(2007)\citenamefont
  {Blumenthal}, \citenamefont {Kaestner}, \citenamefont {Li}, \citenamefont
  {Giblin}, \citenamefont {Janssen}, \citenamefont {Pepper}, \citenamefont
  {Anderson}, \citenamefont {Jones},\ and\ \citenamefont
  {Ritchie}}]{Blumenthal2007}%
  \BibitemOpen
  \bibfield  {author} {\bibinfo {author} {\bibfnamefont {M.~D.}\ \bibnamefont
  {Blumenthal}}, \bibinfo {author} {\bibfnamefont {B.}~\bibnamefont
  {Kaestner}}, \bibinfo {author} {\bibfnamefont {L.}~\bibnamefont {Li}},
  \bibinfo {author} {\bibfnamefont {S.}~\bibnamefont {Giblin}}, \bibinfo
  {author} {\bibfnamefont {T.~J. B.~M.}\ \bibnamefont {Janssen}}, \bibinfo
  {author} {\bibfnamefont {M.}~\bibnamefont {Pepper}}, \bibinfo {author}
  {\bibfnamefont {D.}~\bibnamefont {Anderson}}, \bibinfo {author}
  {\bibfnamefont {G.}~\bibnamefont {Jones}},\ and\ \bibinfo {author}
  {\bibfnamefont {D.~A.}\ \bibnamefont {Ritchie}},\ }\bibfield  {title}
  {\bibinfo {title} {Gigahertz quantized charge pumping},\ }\href
  {https://doi.org/10.1038/nphys582} {\bibfield  {journal} {\bibinfo  {journal}
  {Nature Physics}\ }\textbf {\bibinfo {volume} {3}},\ \bibinfo {pages} {343}
  (\bibinfo {year} {2007})}\BibitemShut {NoStop}%
\bibitem [{\citenamefont {Hermelin}\ \emph {et~al.}(2011)\citenamefont
  {Hermelin}, \citenamefont {Takada}, \citenamefont {Yamamoto}, \citenamefont
  {Tarucha}, \citenamefont {Wieck}, \citenamefont {Saminadayar}, \citenamefont
  {Bäuerle},\ and\ \citenamefont {Meunier}}]{Hermelin2011}%
  \BibitemOpen
  \bibfield  {author} {\bibinfo {author} {\bibfnamefont {S.}~\bibnamefont
  {Hermelin}}, \bibinfo {author} {\bibfnamefont {S.}~\bibnamefont {Takada}},
  \bibinfo {author} {\bibfnamefont {M.}~\bibnamefont {Yamamoto}}, \bibinfo
  {author} {\bibfnamefont {S.}~\bibnamefont {Tarucha}}, \bibinfo {author}
  {\bibfnamefont {A.~D.}\ \bibnamefont {Wieck}}, \bibinfo {author}
  {\bibfnamefont {L.}~\bibnamefont {Saminadayar}}, \bibinfo {author}
  {\bibfnamefont {C.}~\bibnamefont {Bäuerle}},\ and\ \bibinfo {author}
  {\bibfnamefont {T.}~\bibnamefont {Meunier}},\ }\bibfield  {title} {\bibinfo
  {title} {Electrons surfing on a sound wave as a platform for quantum optics
  with flying electrons},\ }\href {https://doi.org/10.1038/nature10416}
  {\bibfield  {journal} {\bibinfo  {journal} {Nature}\ }\textbf {\bibinfo
  {volume} {477}},\ \bibinfo {pages} {435} (\bibinfo {year}
  {2011})}\BibitemShut {NoStop}%
\bibitem [{\citenamefont {McNeil}\ \emph {et~al.}(2011)\citenamefont {McNeil},
  \citenamefont {Kataoka}, \citenamefont {Ford}, \citenamefont {Barnes},
  \citenamefont {Anderson}, \citenamefont {Jones}, \citenamefont {Farrer},\
  and\ \citenamefont {Ritchie}}]{McNeil2011}%
  \BibitemOpen
  \bibfield  {author} {\bibinfo {author} {\bibfnamefont {R.~P.~G.}\
  \bibnamefont {McNeil}}, \bibinfo {author} {\bibfnamefont {M.}~\bibnamefont
  {Kataoka}}, \bibinfo {author} {\bibfnamefont {C.~J.~B.}\ \bibnamefont
  {Ford}}, \bibinfo {author} {\bibfnamefont {C.~H.~W.}\ \bibnamefont {Barnes}},
  \bibinfo {author} {\bibfnamefont {D.}~\bibnamefont {Anderson}}, \bibinfo
  {author} {\bibfnamefont {G.~a.~C.}\ \bibnamefont {Jones}}, \bibinfo {author}
  {\bibfnamefont {I.}~\bibnamefont {Farrer}},\ and\ \bibinfo {author}
  {\bibfnamefont {D.~A.}\ \bibnamefont {Ritchie}},\ }\bibfield  {title}
  {\bibinfo {title} {On-demand single-electron transfer between distant quantum
  dots},\ }\href {https://doi.org/10.1038/nature10444} {\bibfield  {journal}
  {\bibinfo  {journal} {Nature}\ }\textbf {\bibinfo {volume} {477}},\ \bibinfo
  {pages} {439} (\bibinfo {year} {2011})}\BibitemShut {NoStop}%
\bibitem [{\citenamefont {Emary}\ \emph {et~al.}(2019)\citenamefont {Emary},
  \citenamefont {Clark}, \citenamefont {Kataoka},\ and\ \citenamefont
  {Johnson}}]{emary2019energy}%
  \BibitemOpen
  \bibfield  {author} {\bibinfo {author} {\bibfnamefont {C.}~\bibnamefont
  {Emary}}, \bibinfo {author} {\bibfnamefont {L.~A.}\ \bibnamefont {Clark}},
  \bibinfo {author} {\bibfnamefont {M.}~\bibnamefont {Kataoka}},\ and\ \bibinfo
  {author} {\bibfnamefont {N.}~\bibnamefont {Johnson}},\ }\bibfield  {title}
  {\bibinfo {title} {Energy relaxation in hot electron quantum optics via
  acoustic and optical phonon emission},\ }\href@noop {} {\bibfield  {journal}
  {\bibinfo  {journal} {Physical Review B}\ }\textbf {\bibinfo {volume} {99}},\
  \bibinfo {pages} {045306} (\bibinfo {year} {2019})}\BibitemShut {NoStop}%
\bibitem [{\citenamefont {Clark}\ \emph {et~al.}(2020)\citenamefont {Clark},
  \citenamefont {Kataoka},\ and\ \citenamefont {Emary}}]{Clark2020}%
  \BibitemOpen
  \bibfield  {author} {\bibinfo {author} {\bibfnamefont {L.~A.}\ \bibnamefont
  {Clark}}, \bibinfo {author} {\bibfnamefont {M.}~\bibnamefont {Kataoka}},\
  and\ \bibinfo {author} {\bibfnamefont {C.}~\bibnamefont {Emary}},\ }\bibfield
   {title} {\bibinfo {title} {Mitigating decoherence in hot electron
  interferometry},\ }\href {https://doi.org/10.1088/1367-2630/abb9e5}
  {\bibfield  {journal} {\bibinfo  {journal} {New Journal of Physics}\ }\textbf
  {\bibinfo {volume} {22}},\ \bibinfo {pages} {103031} (\bibinfo {year}
  {2020})}\BibitemShut {NoStop}%
\bibitem [{\citenamefont {Yamahata}\ \emph {et~al.}(2016)\citenamefont
  {Yamahata}, \citenamefont {Giblin}, \citenamefont {Kataoka}, \citenamefont
  {Karasawa},\ and\ \citenamefont {Fujiwara}}]{Yamahata2016}%
  \BibitemOpen
  \bibfield  {author} {\bibinfo {author} {\bibfnamefont {G.}~\bibnamefont
  {Yamahata}}, \bibinfo {author} {\bibfnamefont {S.~P.}\ \bibnamefont
  {Giblin}}, \bibinfo {author} {\bibfnamefont {M.}~\bibnamefont {Kataoka}},
  \bibinfo {author} {\bibfnamefont {T.}~\bibnamefont {Karasawa}},\ and\
  \bibinfo {author} {\bibfnamefont {A.}~\bibnamefont {Fujiwara}},\ }\bibfield
  {title} {\bibinfo {title} {Gigahertz single-electron pumping in silicon with
  an accuracy better than 9.2 parts in 107},\ }\href
  {https://doi.org/10.1063/1.4953872} {\bibfield  {journal} {\bibinfo
  {journal} {Applied Physics Letters}\ }\textbf {\bibinfo {volume} {109}},\
  \bibinfo {pages} {013101} (\bibinfo {year} {2016})}\BibitemShut {NoStop}%
\bibitem [{\citenamefont {Levitov}\ \emph {et~al.}(1996)\citenamefont
  {Levitov}, \citenamefont {Lee},\ and\ \citenamefont {Lesovik}}]{Levitov1996}%
  \BibitemOpen
  \bibfield  {author} {\bibinfo {author} {\bibfnamefont {L.~S.}\ \bibnamefont
  {Levitov}}, \bibinfo {author} {\bibfnamefont {H.}~\bibnamefont {Lee}},\ and\
  \bibinfo {author} {\bibfnamefont {G.~B.}\ \bibnamefont {Lesovik}},\
  }\bibfield  {title} {\bibinfo {title} {Electron counting statistics and
  coherent states of electric current},\ }\href
  {https://doi.org/10.1063/1.531672} {\bibfield  {journal} {\bibinfo  {journal}
  {Journal of Mathematical Physics}\ }\textbf {\bibinfo {volume} {37}},\
  \bibinfo {pages} {4845} (\bibinfo {year} {1996})}\BibitemShut {NoStop}%
\bibitem [{\citenamefont {Keeling}\ \emph {et~al.}(2006)\citenamefont
  {Keeling}, \citenamefont {Klich},\ and\ \citenamefont
  {Levitov}}]{Keeling2006}%
  \BibitemOpen
  \bibfield  {author} {\bibinfo {author} {\bibfnamefont {J.}~\bibnamefont
  {Keeling}}, \bibinfo {author} {\bibfnamefont {I.}~\bibnamefont {Klich}},\
  and\ \bibinfo {author} {\bibfnamefont {L.~S.}\ \bibnamefont {Levitov}},\
  }\bibfield  {title} {\bibinfo {title} {Minimal {Excitation} {States} of
  {Electrons} in {One}-{Dimensional} {Wires}},\ }\href
  {https://doi.org/10.1103/PhysRevLett.97.116403} {\bibfield  {journal}
  {\bibinfo  {journal} {Physical Review Letters}\ }\textbf {\bibinfo {volume}
  {97}},\ \bibinfo {pages} {116403} (\bibinfo {year} {2006})}\BibitemShut
  {NoStop}%
\bibitem [{\citenamefont {Ferraro}\ \emph {et~al.}(2014)\citenamefont
  {Ferraro}, \citenamefont {Roussel}, \citenamefont {Cabart}, \citenamefont
  {Thibierge}, \citenamefont {Fève}, \citenamefont {Grenier},\ and\
  \citenamefont {Degiovanni}}]{Ferraro2014}%
  \BibitemOpen
  \bibfield  {author} {\bibinfo {author} {\bibfnamefont {D.}~\bibnamefont
  {Ferraro}}, \bibinfo {author} {\bibfnamefont {B.}~\bibnamefont {Roussel}},
  \bibinfo {author} {\bibfnamefont {C.}~\bibnamefont {Cabart}}, \bibinfo
  {author} {\bibfnamefont {E.}~\bibnamefont {Thibierge}}, \bibinfo {author}
  {\bibfnamefont {G.}~\bibnamefont {Fève}}, \bibinfo {author} {\bibfnamefont
  {C.}~\bibnamefont {Grenier}},\ and\ \bibinfo {author} {\bibfnamefont
  {P.}~\bibnamefont {Degiovanni}},\ }\bibfield  {title} {\bibinfo {title}
  {Real-{Time} {Decoherence} of {Landau} and {Levitov} {Quasiparticles} in
  {Quantum} {Hall} {Edge} {Channels}},\ }\href
  {https://doi.org/10.1103/PhysRevLett.113.166403} {\bibfield  {journal}
  {\bibinfo  {journal} {Physical Review Letters}\ }\textbf {\bibinfo {volume}
  {113}},\ \bibinfo {pages} {166403} (\bibinfo {year} {2014})}\BibitemShut
  {NoStop}%
\bibitem [{\citenamefont {Jullien}\ \emph {et~al.}(2014)\citenamefont
  {Jullien}, \citenamefont {Roulleau}, \citenamefont {Roche}, \citenamefont
  {Cavanna}, \citenamefont {Jin},\ and\ \citenamefont {Glattli}}]{Jullien2014}%
  \BibitemOpen
  \bibfield  {author} {\bibinfo {author} {\bibfnamefont {T.}~\bibnamefont
  {Jullien}}, \bibinfo {author} {\bibfnamefont {P.}~\bibnamefont {Roulleau}},
  \bibinfo {author} {\bibfnamefont {B.}~\bibnamefont {Roche}}, \bibinfo
  {author} {\bibfnamefont {A.}~\bibnamefont {Cavanna}}, \bibinfo {author}
  {\bibfnamefont {Y.}~\bibnamefont {Jin}},\ and\ \bibinfo {author}
  {\bibfnamefont {D.~C.}\ \bibnamefont {Glattli}},\ }\bibfield  {title}
  {\bibinfo {title} {Quantum tomography of an electron},\ }\href
  {https://doi.org/10.1038/nature13821} {\bibfield  {journal} {\bibinfo
  {journal} {Nature}\ }\textbf {\bibinfo {volume} {514}},\ \bibinfo {pages}
  {603} (\bibinfo {year} {2014})}\BibitemShut {NoStop}%
\bibitem [{\citenamefont {Bisognin}\ \emph {et~al.}(2019)\citenamefont
  {Bisognin}, \citenamefont {Marguerite}, \citenamefont {Roussel},
  \citenamefont {Kumar}, \citenamefont {Cabart}, \citenamefont {Chapdelaine},
  \citenamefont {Mohammad-Djafari}, \citenamefont {Berroir}, \citenamefont
  {Bocquillon}, \citenamefont {Plaçais}, \citenamefont {Cavanna},
  \citenamefont {Gennser}, \citenamefont {Jin}, \citenamefont {Degiovanni},\
  and\ \citenamefont {Fève}}]{Bisognin2019}%
  \BibitemOpen
  \bibfield  {author} {\bibinfo {author} {\bibfnamefont {R.}~\bibnamefont
  {Bisognin}}, \bibinfo {author} {\bibfnamefont {A.}~\bibnamefont
  {Marguerite}}, \bibinfo {author} {\bibfnamefont {B.}~\bibnamefont {Roussel}},
  \bibinfo {author} {\bibfnamefont {M.}~\bibnamefont {Kumar}}, \bibinfo
  {author} {\bibfnamefont {C.}~\bibnamefont {Cabart}}, \bibinfo {author}
  {\bibfnamefont {C.}~\bibnamefont {Chapdelaine}}, \bibinfo {author}
  {\bibfnamefont {A.}~\bibnamefont {Mohammad-Djafari}}, \bibinfo {author}
  {\bibfnamefont {J.-M.}\ \bibnamefont {Berroir}}, \bibinfo {author}
  {\bibfnamefont {E.}~\bibnamefont {Bocquillon}}, \bibinfo {author}
  {\bibfnamefont {B.}~\bibnamefont {Plaçais}}, \bibinfo {author}
  {\bibfnamefont {A.}~\bibnamefont {Cavanna}}, \bibinfo {author} {\bibfnamefont
  {U.}~\bibnamefont {Gennser}}, \bibinfo {author} {\bibfnamefont
  {Y.}~\bibnamefont {Jin}}, \bibinfo {author} {\bibfnamefont {P.}~\bibnamefont
  {Degiovanni}},\ and\ \bibinfo {author} {\bibfnamefont {G.}~\bibnamefont
  {Fève}},\ }\bibfield  {title} {\bibinfo {title} {Quantum tomography of
  electrical currents},\ }\href {https://doi.org/10.1038/s41467-019-11369-5}
  {\bibfield  {journal} {\bibinfo  {journal} {Nature Communications}\ }\textbf
  {\bibinfo {volume} {10}},\ \bibinfo {pages} {3379} (\bibinfo {year}
  {2019})}\BibitemShut {NoStop}%
\bibitem [{\citenamefont {Roussely}\ \emph {et~al.}(2018)\citenamefont
  {Roussely}, \citenamefont {Arrighi}, \citenamefont {Georgiou}, \citenamefont
  {Takada}, \citenamefont {Schalk}, \citenamefont {Urdampilleta}, \citenamefont
  {Ludwig}, \citenamefont {Wieck}, \citenamefont {Armagnat}, \citenamefont
  {Kloss}, \citenamefont {Waintal}, \citenamefont {Meunier},\ and\
  \citenamefont {Bäuerle}}]{Roussely2018}%
  \BibitemOpen
  \bibfield  {author} {\bibinfo {author} {\bibfnamefont {G.}~\bibnamefont
  {Roussely}}, \bibinfo {author} {\bibfnamefont {E.}~\bibnamefont {Arrighi}},
  \bibinfo {author} {\bibfnamefont {G.}~\bibnamefont {Georgiou}}, \bibinfo
  {author} {\bibfnamefont {S.}~\bibnamefont {Takada}}, \bibinfo {author}
  {\bibfnamefont {M.}~\bibnamefont {Schalk}}, \bibinfo {author} {\bibfnamefont
  {M.}~\bibnamefont {Urdampilleta}}, \bibinfo {author} {\bibfnamefont
  {A.}~\bibnamefont {Ludwig}}, \bibinfo {author} {\bibfnamefont {A.~D.}\
  \bibnamefont {Wieck}}, \bibinfo {author} {\bibfnamefont {P.}~\bibnamefont
  {Armagnat}}, \bibinfo {author} {\bibfnamefont {T.}~\bibnamefont {Kloss}},
  \bibinfo {author} {\bibfnamefont {X.}~\bibnamefont {Waintal}}, \bibinfo
  {author} {\bibfnamefont {T.}~\bibnamefont {Meunier}},\ and\ \bibinfo {author}
  {\bibfnamefont {C.}~\bibnamefont {Bäuerle}},\ }\bibfield  {title} {\bibinfo
  {title} {Unveiling the bosonic nature of an ultrashort few-electron pulse},\
  }\href {https://doi.org/10.1038/s41467-018-05203-7} {\bibfield  {journal}
  {\bibinfo  {journal} {Nature Communications}\ }\textbf {\bibinfo {volume}
  {9}},\ \bibinfo {pages} {2811} (\bibinfo {year} {2018})}\BibitemShut
  {NoStop}%
\bibitem [{\citenamefont {Vyshnevyy}\ \emph {et~al.}(2013)\citenamefont
  {Vyshnevyy}, \citenamefont {Lebedev}, \citenamefont {Lesovik},\ and\
  \citenamefont {Blatter}}]{Vyshnevyy2013}%
  \BibitemOpen
  \bibfield  {author} {\bibinfo {author} {\bibfnamefont {A.~A.}\ \bibnamefont
  {Vyshnevyy}}, \bibinfo {author} {\bibfnamefont {A.~V.}\ \bibnamefont
  {Lebedev}}, \bibinfo {author} {\bibfnamefont {G.~B.}\ \bibnamefont
  {Lesovik}},\ and\ \bibinfo {author} {\bibfnamefont {G.}~\bibnamefont
  {Blatter}},\ }\bibfield  {title} {\bibinfo {title} {Two-particle entanglement
  in capacitively coupled mach-zehnder interferometers},\ }\href
  {https://doi.org/10.1103/physrevb.87.165302} {\bibfield  {journal} {\bibinfo
  {journal} {Physical Review B}\ }\textbf {\bibinfo {volume} {87}},\ \bibinfo
  {pages} {165302} (\bibinfo {year} {2013})}\BibitemShut {NoStop}%
\bibitem [{\citenamefont {Dasenbrook}\ \emph {et~al.}(2016)\citenamefont
  {Dasenbrook}, \citenamefont {Bowles}, \citenamefont {Brask}, \citenamefont
  {Hofer}, \citenamefont {Flindt},\ and\ \citenamefont
  {Brunner}}]{Dasenbrook2016a}%
  \BibitemOpen
  \bibfield  {author} {\bibinfo {author} {\bibfnamefont {D.}~\bibnamefont
  {Dasenbrook}}, \bibinfo {author} {\bibfnamefont {J.}~\bibnamefont {Bowles}},
  \bibinfo {author} {\bibfnamefont {J.~B.}\ \bibnamefont {Brask}}, \bibinfo
  {author} {\bibfnamefont {P.~P.}\ \bibnamefont {Hofer}}, \bibinfo {author}
  {\bibfnamefont {C.}~\bibnamefont {Flindt}},\ and\ \bibinfo {author}
  {\bibfnamefont {N.}~\bibnamefont {Brunner}},\ }\bibfield  {title} {\bibinfo
  {title} {Single-electron entanglement and nonlocality},\ }\href
  {https://doi.org/10.1088/1367-2630/18/4/043036} {\bibfield  {journal}
  {\bibinfo  {journal} {New Journal of Physics}\ }\textbf {\bibinfo {volume}
  {18}},\ \bibinfo {pages} {043036} (\bibinfo {year} {2016})}\BibitemShut
  {NoStop}%
\bibitem [{\citenamefont {Moskalets}(2016)}]{Moskalets2016}%
  \BibitemOpen
  \bibfield  {author} {\bibinfo {author} {\bibfnamefont {M.}~\bibnamefont
  {Moskalets}},\ }\bibfield  {title} {\bibinfo {title} {Fractionally charged
  zero-energy single-particle excitations in a driven fermi sea},\ }\href
  {https://doi.org/10.1103/PhysRevLett.117.046801} {\bibfield  {journal}
  {\bibinfo  {journal} {Physical Review Letters}\ }\textbf {\bibinfo {volume}
  {117}},\ \bibinfo {pages} {046801} (\bibinfo {year} {2016})}\BibitemShut
  {NoStop}%
\bibitem [{\citenamefont {Ronetti}\ \emph {et~al.}(2018)\citenamefont
  {Ronetti}, \citenamefont {Vannucci}, \citenamefont {Ferraro}, \citenamefont
  {Jonckheere}, \citenamefont {Rech}, \citenamefont {Martin},\ and\
  \citenamefont {Sassetti}}]{Ronetti2018}%
  \BibitemOpen
  \bibfield  {author} {\bibinfo {author} {\bibfnamefont {F.}~\bibnamefont
  {Ronetti}}, \bibinfo {author} {\bibfnamefont {L.}~\bibnamefont {Vannucci}},
  \bibinfo {author} {\bibfnamefont {D.}~\bibnamefont {Ferraro}}, \bibinfo
  {author} {\bibfnamefont {T.}~\bibnamefont {Jonckheere}}, \bibinfo {author}
  {\bibfnamefont {J.}~\bibnamefont {Rech}}, \bibinfo {author} {\bibfnamefont
  {T.}~\bibnamefont {Martin}},\ and\ \bibinfo {author} {\bibfnamefont
  {M.}~\bibnamefont {Sassetti}},\ }\bibfield  {title} {\bibinfo {title}
  {Crystallization of levitons in the fractional quantum {Hall} regime},\
  }\href {https://doi.org/10.1103/PhysRevB.98.075401} {\bibfield  {journal}
  {\bibinfo  {journal} {Physical Review B}\ }\textbf {\bibinfo {volume} {98}},\
  \bibinfo {pages} {075401} (\bibinfo {year} {2018})}\BibitemShut {NoStop}%
\bibitem [{\citenamefont {Gaury}\ and\ \citenamefont
  {Waintal}(2014)}]{Gaury2014}%
  \BibitemOpen
  \bibfield  {author} {\bibinfo {author} {\bibfnamefont {B.}~\bibnamefont
  {Gaury}}\ and\ \bibinfo {author} {\bibfnamefont {X.}~\bibnamefont
  {Waintal}},\ }\bibfield  {title} {\bibinfo {title} {Dynamical control of
  interference using voltage pulses in the quantum regime},\ }\href
  {https://doi.org/10.1038/ncomms4844} {\bibfield  {journal} {\bibinfo
  {journal} {Nature Communications}\ }\textbf {\bibinfo {volume} {5}},\
  \bibinfo {pages} {3844} (\bibinfo {year} {2014})}\BibitemShut {NoStop}%
\bibitem [{\citenamefont {Burset}\ \emph {et~al.}(2019)\citenamefont {Burset},
  \citenamefont {Kotilahti}, \citenamefont {Moskalets},\ and\ \citenamefont
  {Flindt}}]{Burset2019}%
  \BibitemOpen
  \bibfield  {author} {\bibinfo {author} {\bibfnamefont {P.}~\bibnamefont
  {Burset}}, \bibinfo {author} {\bibfnamefont {J.}~\bibnamefont {Kotilahti}},
  \bibinfo {author} {\bibfnamefont {M.}~\bibnamefont {Moskalets}},\ and\
  \bibinfo {author} {\bibfnamefont {C.}~\bibnamefont {Flindt}},\ }\bibfield
  {title} {\bibinfo {title} {Time-domain spectroscopy of mesoscopic conductors
  using voltage pulses},\ }\href {https://doi.org/10.1002/qute.201900014}
  {\bibfield  {journal} {\bibinfo  {journal} {Advanced Quantum Technologies}\
  }\textbf {\bibinfo {volume} {2}},\ \bibinfo {pages} {1900014} (\bibinfo
  {year} {2019})}\BibitemShut {NoStop}%
\bibitem [{\citenamefont {Udem}\ \emph {et~al.}(2002)\citenamefont {Udem},
  \citenamefont {Holzwarth},\ and\ \citenamefont {Hänsch}}]{Udem2002}%
  \BibitemOpen
  \bibfield  {author} {\bibinfo {author} {\bibfnamefont {T.}~\bibnamefont
  {Udem}}, \bibinfo {author} {\bibfnamefont {R.}~\bibnamefont {Holzwarth}},\
  and\ \bibinfo {author} {\bibfnamefont {T.~W.}\ \bibnamefont {Hänsch}},\
  }\bibfield  {title} {\bibinfo {title} {Optical frequency metrology},\ }\href
  {https://doi.org/10.1038/416233a} {\bibfield  {journal} {\bibinfo  {journal}
  {Nature}\ }\textbf {\bibinfo {volume} {416}},\ \bibinfo {pages} {233}
  (\bibinfo {year} {2002})}\BibitemShut {NoStop}%
\bibitem [{\citenamefont {Moskalets}\ and\ \citenamefont
  {Büttiker}(2002)}]{Moskalets2002}%
  \BibitemOpen
  \bibfield  {author} {\bibinfo {author} {\bibfnamefont {M.}~\bibnamefont
  {Moskalets}}\ and\ \bibinfo {author} {\bibfnamefont {M.}~\bibnamefont
  {Büttiker}},\ }\bibfield  {title} {\bibinfo {title} {Floquet scattering
  theory of quantum pumps},\ }\href
  {https://doi.org/10.1103/PhysRevB.66.205320} {\bibfield  {journal} {\bibinfo
  {journal} {Physical Review B}\ }\textbf {\bibinfo {volume} {66}},\ \bibinfo
  {pages} {205320} (\bibinfo {year} {2002})}\BibitemShut {NoStop}%
\bibitem [{\citenamefont {Dubois}\ \emph
  {et~al.}(2013{\natexlab{b}})\citenamefont {Dubois}, \citenamefont {Jullien},
  \citenamefont {Grenier}, \citenamefont {Degiovanni}, \citenamefont
  {Roulleau},\ and\ \citenamefont {Glattli}}]{Dubois2013a}%
  \BibitemOpen
  \bibfield  {author} {\bibinfo {author} {\bibfnamefont {J.}~\bibnamefont
  {Dubois}}, \bibinfo {author} {\bibfnamefont {T.}~\bibnamefont {Jullien}},
  \bibinfo {author} {\bibfnamefont {C.}~\bibnamefont {Grenier}}, \bibinfo
  {author} {\bibfnamefont {P.}~\bibnamefont {Degiovanni}}, \bibinfo {author}
  {\bibfnamefont {P.}~\bibnamefont {Roulleau}},\ and\ \bibinfo {author}
  {\bibfnamefont {D.~C.}\ \bibnamefont {Glattli}},\ }\bibfield  {title}
  {\bibinfo {title} {Integer and fractional charge {Lorentzian} voltage pulses
  analyzed in the framework of photon-assisted shot noise},\ }\href
  {https://doi.org/10.1103/PhysRevB.88.085301} {\bibfield  {journal} {\bibinfo
  {journal} {Physical Review B}\ }\textbf {\bibinfo {volume} {88}},\ \bibinfo
  {pages} {085301} (\bibinfo {year} {2013}{\natexlab{b}})}\BibitemShut
  {NoStop}%
\bibitem [{\citenamefont {Auston}(1975)}]{Auston1975}%
  \BibitemOpen
  \bibfield  {author} {\bibinfo {author} {\bibfnamefont {D.~H.}\ \bibnamefont
  {Auston}},\ }\bibfield  {title} {\bibinfo {title} {Picosecond optoelectronic
  switching and gating in silicon},\ }\href {https://doi.org/10.1063/1.88079}
  {\bibfield  {journal} {\bibinfo  {journal} {Applied Physics Letters}\
  }\textbf {\bibinfo {volume} {26}},\ \bibinfo {pages} {101} (\bibinfo {year}
  {1975})}\BibitemShut {NoStop}%
\bibitem [{\citenamefont {Georgiou}\ \emph {et~al.}(2020)\citenamefont
  {Georgiou}, \citenamefont {Geffroy}, \citenamefont {Bäuerle},\ and\
  \citenamefont {Roux}}]{Georgiou2020}%
  \BibitemOpen
  \bibfield  {author} {\bibinfo {author} {\bibfnamefont {G.}~\bibnamefont
  {Georgiou}}, \bibinfo {author} {\bibfnamefont {C.}~\bibnamefont {Geffroy}},
  \bibinfo {author} {\bibfnamefont {C.}~\bibnamefont {Bäuerle}},\ and\
  \bibinfo {author} {\bibfnamefont {J.-F.}\ \bibnamefont {Roux}},\ }\bibfield
  {title} {\bibinfo {title} {Efficient three-dimensional
  photonic{\textendash}plasmonic photoconductive switches for picosecond {THz}
  pulses},\ }\href {https://doi.org/10.1021/acsphotonics.0c00044} {\bibfield
  {journal} {\bibinfo  {journal} {{ACS} Photonics}\ }\textbf {\bibinfo {volume}
  {7}},\ \bibinfo {pages} {1444} (\bibinfo {year} {2020})}\BibitemShut
  {NoStop}%
\end{thebibliography}%





\end{document}